\documentclass[10pt,aps,superscriptaddress,prc,eprint,nofootinbib,showkeys]{revtex4-2}
\pdfoutput=1
\usepackage{graphicx}
\usepackage{dcolumn}
\usepackage{bm}
\usepackage{amsmath,amsfonts,amsthm,amssymb}
\usepackage{graphics}
\usepackage{cancel}
\usepackage{multirow}
\usepackage{longtable}
\usepackage{color}
\usepackage[normalem]{ulem} 
\usepackage{physics}
\usepackage{comment}
\usepackage{amssymb}
\usepackage{graphicx}
\usepackage{amsmath}
\usepackage{subfigure}
\usepackage{bbold}
\usepackage{yfonts}
\usepackage[colorlinks=true,linktocpage=true,linkcolor=blue,citecolor=blue]{hyperref}
\usepackage{placeins}
\usepackage{bm} 
\usepackage{nicefrac}
\usepackage{slashed}
\usepackage[dvipsnames]{xcolor}
\usepackage{marginnote}

\newcommand{\ab}[1]{\left\langle#1\right\rangle}

\newcommand{\intp}{\int_{\textbf{p}}}
\newcommand{\ep}{E_{\textbf{p}}}

\newcommand{\taur}{\tau_{\rm R}}
\newcommand{\faz}{f_a^{\,0}}
\newcommand{\afaz}{\Bar{f}_a^{\,0}}
\newcommand{\tfaz}{\Tilde{f}_a^{\,0}}
\newcommand{\tafaz}{\Tilde{\Bar{f}}_a^{\,0}}
\newcommand{\dfa}{\delta f_a}
\newcommand{\dafa}{\delta \Bar{f}_a}

\begin{document}
	
	\preprint{APS/123-QED}

    \title{Diffusion of multiple conserved charges from entropy production}
    
	\author{Samapan Bhadury}
	\email{samapanb@iiserbpr.ac.in}
	\affiliation{Department of Physical Sciences,\\ Indian Institute of Science Education and Research Berhampur, Laudigam–760003, Dist.–Ganjam, Odisha, India}
	
	\author{Arpan Das}
	\email{arpan.das@pilani.bits-pilani.ac.in}
	\affiliation{Department of Physics, Birla Institute of Technology and Science Pilani, Pilani Campus, Pilani, Rajasthan-333031, India}
	
	\author{Sandeep Chatterjee}
	\email{sandeep@iiserbpr.ac.in}
	\affiliation{Department of Physical Sciences,\\ Indian Institute of Science Education and Research Berhampur, Laudigam–760003, Dist.–Ganjam, Odisha, India}
    
	\author{Hiranmaya Mishra}
	\email{hiranmaya@niser.ac.in}
	\affiliation{School of Physical Sciences, National Institute of Science Education and Research, An OCC of Homi Bhabha National Institute, Jatni-752050, India}
	\affiliation{Institute of Physics, Sachivalaya Marg, Bhubaneswar-751005, India}

    \begin{abstract}
        We derive dissipative relativistic hydrodynamic equations in the presence of multiple conserved charges, i.e., baryon number ($B$), electric charge ($Q$), and strangeness ($S$), using the Chapman-Enskog (CE) method within the kinetic theory approach. The relativistic Boltzmann equation is solved within the relaxation-time approximation with a momentum-independent relaxation time in the collision term. We derive both first-order (Navier-Stokes limit) and second-order dissipative hydrodynamic equations. Within the kinetic theory framework, using the Boltzmann's H-theorem, and by demanding that for a dissipative system, the entropy must be produced, we find different transport coefficients at the first-order and second-order gradient expansion of the out-of-equilibrium distribution function around the local equilibrium. Apart from the well-known transport coefficients, the shear ($\eta$) and the bulk ($\zeta$) viscosities , we also find the diffusion matrix elements ($\kappa_{qq^{\prime}}$) for the conserved charges $B$, $Q$ and $S$. The diffusion matrix elements ($\kappa_{qq^{\prime}}$) are important to model the multi-component diffusion dynamics sourced by inhomogeneous baryon stopping in the initial state of heavy-ion collisions. We estimate the temperature ($T$) and chemical potential dependence of diagonal and off-diagonal elements of the diffusion matrix elements for the (2+1) flavor quark-gluon plasma. We further estimate the ratio $\kappa_{qq^{\prime}}T/\eta$ for a wide range of temperature and chemical potentials to show the relative importance of the diffusion matrix elements compared to other transport coefficients. 
    \end{abstract}
         
    \date{\today}

\maketitle

\section{Introduction}
Relativistic kinetic theory models and relativistic hydrodynamic frameworks allow us to model the evolution of QCD plasma produced in heavy-ion collision (HIC) experiments~\cite{Kolb:2000fha, Ollitrault:2007du, Schenke:2010rr, Romatschke:2009im, Huovinen:2011xc, Jaiswal:2016hex, Florkowski:2017olj, Romatschke:2017ejr, Fotakis:2019nbq, Bhalerao:2020ulk}. These models have been successfully applied to describe the evolution of the fireball created at ultra-relativistic heavy-ion collisions at the Relativistic Heavy Ion Collider (RHIC) and at the Large Hadron Collider (LHC). Most studies in this field so far have primarily examined the high-temperature, low-baryon-chemical-potential region of the QCD phase diagram \cite{Halasz:1998qr, Aoki:2006we, Borsanyi:2010bp, Bazavov:2011nk}. It is of paramount interest to study the QCD phase structure and the physics of strongly interacting matter at finite baryon density, relevant to relativistic nuclear collisions at lower collision energies~\cite{Fukushima:2013rx,Fukushima:2010bq}.
This research direction has been pursued actively in several experimental programmes, including the NA61/SHINE experiment at the SPS, the Beam Energy Scan programme at RHIC, and there are plans to perform similar studies in the upcoming Compressed Baryonic Matter experiment at FAIR in Darmstadt, Germany. At finite baryon density, the QCD phase diagram may feature a QCD critical endpoint, where the theoretically conjectured first-order phase transition line ends~\cite{Fukushima:2013rx, Stephanov:1998dy, Stephanov:1999zu}. The search for the QCD critical point is among the central physics objectives of low energy heavy-ion collision  facilities~\cite{STAR:2010vob, Mohanty:2011nm, Mitchell:2012mx, Rajagopal:2019xwg, Martinez:2019bsn, Dore:2020jye, Grefa:2022sav}.

Considering the success of relativistic kinetic theory models and relativistic hydrodynamic frameworks in explaining the space-time evolution of the QCD plasma at ultra relativistic collisions \cite{Jeon:1995zm, Denicol:2014mca, Jaiswal:2014isa, Florkowski:2017ruc, Mykhaylova:2019wci}, development of these models for low-energy collisions has also gained a lot of attention \cite{Plumberg:2024leb, Jaiswal:2015mxa, Denicol:2018wdp}. As compared to ultra-relativistic collisions, in the low-energy heavy-ion collision experiments, baryon-rich QCD plasma in produced. Hence, the low energy collisions may provide an opportunity to investigate the evolution of the QCD medium in the vicinity of the QCD critical point or QCD critical region. Correlations and fluctuations of conserved charges carry the information about the dynamics of the medium near the critical point. In this context, often one studies the net-baryon number fluctuation to explore the scaling properties of fluctuations~\cite{Stephanov:1998dy, Stephanov:1999zu, Stephanov:2004wx, Kapusta:2011gt, Young:2014pka}. QCD equation of state incorporating the effect of the critical point have been discussed in Refs.~\cite{Nonaka:2004pg, Paech:2005cx, Herold:2013bi, Mukherjee:2015swa}. Effect of baryon diffusion has also been discussed within the hydrodynamic simulations~\cite{Schenke:2010nt, Denicol:2018wdp}. Often, the net baryon number is considered as the only conserved charge hydrodynamic frameworks. However, if one considers that all the light quark flavors, i.e. 
up ($u$), down ($d$), and strange ($s$) thermalize in a strongly interacting medium, then the net baryon ($B$), electric charge ($Q$), strangeness ($S$), and their cross-correlations needs to be included in the hydrodynamic framework~\cite{Fotakis:2019nbq, Denicol:2018wdp, Du:2019obx, Schafer:2021csj, Du:2021zqz, Almaalol:2022pjc, Monnai:2021kgu, Monnai:2019hkn}. Such frameworks are commonly known as the $BQS$ - hydrodynamic framework for the QCD medium at the finite baryon density~\cite{Plumberg:2024leb, Denicol:2018wdp, Du:2019obx, Fotakis:2019nbq}.

It may be noted here that the fluctuations of the above conserved charges play an important role in the search for the critical point \cite{Asakawa:2015ybt}. It has been argued that the fluctuations of conserved charges can be a possible signature of quark-hadron phase transition and QGP formation. Further, due to the rapid expansion of the fireball, the fluctuations originated at the QGP phase may survive until the freeze-out and can be used as a signal of QGP formation in the early stages of the collision~\cite{Asakawa:2000wh}. In the context of these conserved charge fluctuations, diffusion plays an important role, as the time evolution of the conserved charges can be caused by diffusion processes. Indeed, the fluctuations of the baryon number $(B)$, electric charge $(Q)$, and strangeness $(S)$ can affect the strangeness flow, and diffusion of these conserved charges can affect the rapidity dependence of the charge distribution~\cite{Fotakis:2019nbq, Martinez:2019jbu, Carzon:2019qja}. BQS diffusion can non-trivially affect the QCD dynamics at the critical point~\cite{Rougemont:2015ona, Grefa:2022sav, Rougemont:2017tlu, Martinez:2019bsn, Rajagopal:2019xwg, Dore:2020jye, Dore:2022qyz}. Interestingly, in the presence of multiple conserved charges, the diffusion of one conserved charge gets nontrivial contributions from other charges. These nontrivial contributions are encoded in the diffusion matrix elements. Due to a non trivial diffusion matrix, the diffusion current of each conserved charge will no longer depend solely on the gradient of that specific charge. Since the gradients of every single charge density can generate a diffusion current of any other charge, the diffusion currents of the conserved charges must be coupled \cite{Hu:2022vph, Greif:2017byw}. The presence of these cross-diffusion coefficients brings novelty to the hydrodynamic framework at finite baryon density. Diffusion dynamics of multiple conserved charges and theoretical development of multi-component dissipative hydrodynamics has been discussed in Refs.~\cite{Greif:2017byw, Fotakis:2019nbq, Rose:2020sjv, Fotakis:2021diq, Fotakis:2022usk, Hu:2022vph, Fotakis:2024hmz}. Second-order evolution equations have been obtained for multi-flavors in Ref.~\cite{Harutyunyan:2023nvt} from entropy flow under the framework of relativistic hydrodynamics, whereas in Ref.~\cite{Dey:2024hhc}, the authors used non-equilibrium statistical operator method, generalizing the method devised by Zubarev to multi-species system and also wrote down the Kubo relations. There has also been some studies on a system with multiple charges from the point of view of holography \cite{Jain:2009pw, Gladden:2025glw}. Understanding the second-order evolution equations are of importance to ensure stable and causal formulation and simulation of the strongly interacting medium in HIC experiments \cite{Almaalol:2022pjc, Cordeiro:2025mtg}. In Refs.~\cite{Hu:2022vph, Fotakis:2022usk} authors have obtained multi-component relativistic dissipative fluid dynamics within the framework of kinetic theory approach using the moment-expansion method~\cite{Denicol:2012cn}. The influence of magnetic field for a two-component system was also investigated recently in Ref.~\cite{Kushwah:2025jsb}. In the moment method, the out-of-equilibrium distribution function is expressed as a Taylor series around the equilibrium part in powers of four-momenta. Deriving hydrodynamic equations by solving the kinetic theory equations with exact collision integrals can be a non-trivial task. In such situations, one may avoid the explicit evaluation of the collision term by employing Chapman-Enskog-like (CE) expansion of the distribution function in the relaxation time approximation~\cite{Jaiswal:2013npa}.

In this work, we use Chapman-Enskog-like expansion and the relaxation time approximation (RTA) to obtain multi-component fluid dynamics by determining the entropy production for a mix of massive and massless particles. We obtain the second-order evolution equations of the dissipative currents, present in the medium. This method has the advantage of producing only those physical transport coefficients that are dissipative in nature. Additionally, unlike the some previous studies \cite{Fotakis:2019nbq}, the diagonal components of the diffusion matrix are manifestly positive. Here, the out-of-equilibrium distribution function is expanded around the equilibrium distribution, ordered in powers of space-time gradients. This framework provides a direct correspondence between the gradient order of the distribution function and the hydrodynamic gradient expansion. Moreover, within the RTA, the CE method also enables the determination of the transport coefficients, including the diffusion matrix elements, that appear in the hydrodynamic equations. We specifically focus on the components of the diffusion matrix, their importance relative to the shear viscosity and also study some their properties in some limiting cases. In our calculation, we do not use any quasi-particle picture for different partons.

The rest of the manuscript is organized in the following manner. In Sec.~\ref{secII} we introduce the CE method and the relaxation time approximation. In Secs.~\ref{ssec:FOTP} and \ref{ssec:SOTP} we obtain the transport coefficients in the first-order and second-order hydrodynamic theory. In Sec.~\ref{sec:R&D} we present important results, and present numerical estimates of various transport coefficients, importantly, the cross-diffusion coefficients of quark gluon plasma. Finally, in Sec.~\ref{sec:C&O} we draw conclusions, and present possible future directions.

\textit{Notation and conventions:} Throughout the article, we have assumed natural units ($\hbar = c = k_{\rm B} = 1$). We will be considering a flat spacetime with the metric tensor given by, $g_{\mu\nu} = {\rm diag}(1,-1,-1,-1)$. We have also used the three index fully anti-symmetric Levi-Civita symbol, $\varepsilon_{ijk}$. For brevity, throughout the article, we will use $\intp \equiv \int \frac{g_a d^3 \textbf{p}}{(2\pi)^3 \ep}$, where, $g_a$ is degeneracy of the $a$-th species and, $\ep \equiv \sqrt{\textbf{p}^2 + m^2}$ is the particle energy. For the mixture of particle species, while the particle's momenta, energy and mass should carry an index representing the species, we have suppressed such species indices to keep the notations under control. However, we will add species indices to the phase space distribution functions as, $f_a$ ($\Bar{f}_a$) for the particles (anti-particles) but suppress the dependence on the phase-space variables. Also note that, we may place the labels like $a$ and $q$ (which are not to be confused with any Lorentz indices or three-vector indices) in both superscript and subscript of some variables as per the convenience of notation. On multiple occasions in the article, we will have to sum over the particle species and/or, the conserved charges. For the conserved charges the notation, $\sum_q (\cdots)$ should be understood as sum over the charges $B,Q,S$. On the other hand, we will use two types of summations for the particle species - (i) $\sum_a (\cdots)$ will be used when summing over the fermionic particles only i.e., $u, d, s$ and, (ii) ${\sum_a}{\!\!'} (\cdots)$ will be used when summing over all particles of the systems i.e., $u, d, s, g$. The inner product between two four vectors $A^{\mu}$, and $B^{\mu}$ is denoted as, $A\cdot B\equiv A^{\mu}B_{\mu}$. 

%-----------------------------------
\section{Relativistic Hydrodynamics}
\label{secII}
%-----------------------------------

To develop a theory of relativistic hydrodynamics of a mixture of quarks and gluons, we may assume the system under consideration have three conserved particle species which will be called quarks ($a = u, d, s$) each carrying three possible types of charges ($q = B, Q, S$). Additionally the system may consist of massless, chargeless, bosonic particles,which we will call gluons ($g$). The gluons are not conserved. The conservation of individual particle species ensures the conservation of charges by construction. Thus we can write the conserved currents as,
\begin{subequations}
    \begin{align}
        %1
        N_a^\mu &= n_a u^\mu + n_a^\mu = \intp p^\mu \left(f_a - \Bar{f}_a\right), \hspace{4.6cm} (a = u,d,s) \label{N_a^m-def} \\
        %2
        T^{\mu\nu} &= \varepsilon u^\mu u^\nu - \left(P + \Pi \right) \Delta^{\mu\nu} + \pi^{\mu\nu} = {\sum_a}' \intp p^\mu p^\nu \left(f_a + \Bar{f}_a\right),
        \label{T^mn-def}
    \end{align}
\end{subequations}
where, $n_a$ and $n_a^\mu$ are the net particle number density and net particle diffusion current of the $a$-th species respectively, $\varepsilon, P, \Pi$ and, $\pi^{\mu\nu}$ are the total energy density, isotropic pressure, bulk viscous pressure and shear pressure of the system respectively. The tensor decomposition for $N_a^\mu$ and $T^{\mu\nu}$ is written under the Landau-Lifshitz definition of the fluid four velocity, $u^\mu$ i.e., $u_\nu T^{\mu\nu} = \varepsilon u^\mu$. Additionally, we have also imposed the Landau-Lifshitz matching conditions, which makes $n_a$, $\varepsilon$, equilibrium quantities. The projection operator $\Delta^{\mu\nu} = g^{\mu\nu} - u^\mu u^\nu$ is orthogonal to fluid four-velocity. The phase-space distribution functions for particles and anti-particles of $a$-th species are denoted by, $\faz$ and $\afaz$ respectively. We define the $q$-type conserved charge current as,
\begin{align}
    J_q^\mu = \sum_a q_a N_a^\mu = n_q u^\mu + n_q^\mu = \sum_a q_a \intp p^\mu \left(f_a - \Bar{f}_a\right), \hspace{1cm} (q = B,Q,S)\label{J_q^m-def}
\end{align}
where $q_a$ is the $q$-type charge of the $a$-th particle species, $n_q$ and, $n_q^\mu$ are the charge density and charge diffusion currents of $q$-type. We note the conservation laws of the system are given by,
\begin{align}
    \partial_\mu N_a^\mu = 0 \implies \partial_\mu J_q^\mu = 0,
    \quad{\rm and,}\quad
    \partial_\mu T^{\mu\nu} = 0. \hspace{2cm} (a = u,d,s~~{\rm and},~~q = B,Q,S) \label{consv.laws}
\end{align}
Substituting Eqs.~\eqref{N_a^m-def} and \eqref{T^mn-def} into Eq.~\eqref{consv.laws} and taking the projections along and orthogonal to the fluid four-velocity, we obtain \cite{DeGroot:1980dk, Romatschke:2007mq, Jaiswal:2015mxa},
\begin{subequations}
    \begin{align}
        %1
        \Dot{n}_a + n_a\, \theta + \left(\partial\cdot n_a\right) &= 0, \hspace{2cm} (a = u,d,s) \label{heq1a}\\
        %2
        \Dot{n}_q + n_q\, \theta + \left(\partial\cdot n_q\right) &= 0, \hspace{2cm} (q = B,Q,S) \label{heq1b}\\
        %3
        \Dot{\varepsilon} + \left(\varepsilon + P\right) \theta + \Pi \theta - \pi^{\mu\nu} \sigma_{\mu\nu} &= 0, \label{heq2}\\
        %4
        \left(\varepsilon + P\right) \Dot{u}^\mu - \left(\nabla^\mu P\right) + \Delta^\mu_\alpha \partial_\beta \pi^{\alpha\beta} &= 0, \label{heq3}
    \end{align}
\end{subequations}
where, the co-moving derivative is defined as, $\dot{A}=u^{\mu}\partial_{\mu}A$, $\theta \equiv \partial_\mu u^\mu$ is the expansion scalar, and, $\sigma_{\mu\nu} \equiv \Delta_{\mu\nu}^{\alpha\beta} \left(\partial_\alpha u_\beta\right)$ is the velocity shear stress tensor with $\Delta_{\mu\nu}^{\alpha\beta} \equiv (1/2) (\Delta_{~\mu}^\alpha \Delta_{~\nu}^\beta + \Delta_{~\nu}^\alpha \Delta_{~\mu}^\beta) - (1/3) \Delta_{\mu\nu} \Delta^{\alpha\beta}$ being the doubly symmetric traceless rank-4 projection operator. Derivative normal to the fluid flow is defined as, $\nabla^{\mu}\equiv \Delta^{\mu\nu}\partial_{\nu}$. We can define the variables used in Eqs.~\eqref{heq1a}-\eqref{heq3} in terms of the phase-space distribution functions, which can be split into equilibrium and out-of-equilibrium parts, following the Chapman-Enskog like expansion as, $f_a = \faz + \dfa$ (and similarly for anti-particles). Then we can write,
\begin{subequations}
    \begin{align}
        %1
        n_q &= \sum_{a} q_a n_a = \sum_a q_a \intp \ep \left(\faz - \afaz\right) = \sum_a q_a I_{10}^{a,-}, \hspace{2cm} (q = B,Q,S) \label{n_q-def} \\
        %2
        \varepsilon &= {\sum_a}' q_a \intp \ep^2 \left(\faz + \afaz\right) = {\sum_a}' I_{20}^{a,+} = I_{20}^+, \label{ve-def} \\
        %3
        P &= - \frac{1}{3} {\sum_{a}}' \intp \left(p\cdot\Delta\cdot p\right) \left(\faz + \afaz\right) = - {\sum_a}' I_{21}^{a,+} = - I_{21}^{+}, \label{P-def} \\
        %4
        \Pi &= - \frac{1}{3} {\sum_{a}}' \intp \left(p\cdot\Delta\cdot p\right) \left(\dfa + \dafa\right), \label{Pi-def} \\
        %5
        n_q^\mu &= \sum_{a} q_a n_a^\mu = \sum_a q_a \intp p^{\ab{\mu}} \left(\dfa - \dafa\right), \hspace{3.7cm} (q = B,Q,S) \label{n_q^m-def} \\
        %6
        \pi^{\mu\nu} &= {\sum_{a}}' \intp p^{\langle\mu} p^{\nu\rangle} \left(\dfa + \dafa\right), \label{pi^mn-def}
    \end{align}
\end{subequations}
where, we have expressed the equilibrium variables in terms of the thermodynamic integrals defined as,
\begin{align}
    I_{nq}^{a,\pm} \equiv \frac{1}{(2q+1)!!} \intp \ep^{n-2q} \left(p\cdot\Delta\cdot p\right)^q \left(\faz \pm \afaz \right), \label{Inq}
\end{align}
and we have also used the definition, $I_{nq}^{\pm} \equiv \sum_a I_{nq}^{a,\pm}$ to simplify the notations, here and from this point onward whenever possible. In Eqs.~\eqref{n_q^m-def} and \eqref{pi^mn-def} we have introduced the notations, $A^{\langle\mu_1\cdots\mu_n\rangle} \equiv \Delta^{\mu_1\cdots\mu_n}_{\alpha_1\cdots\alpha_n} A^{\alpha_1\cdots \alpha_n}$, where $\Delta^{\mu_1\cdots\mu_n}_{\alpha_1\cdots\alpha_n}$ is a projection operator that is symmetric and traceless in individual type of indices (either $\mu_i$ or $\alpha_i$), whose properties\footnote{Note that the traceless property is not meaningful for $n=1$ case and hence holds true for $n\geq2$ cases.} can be found in Refs.~\cite{DeGroot:1980dk, Denicol:2012cn}. 
Using these projection operators, we may construct a set of irreducible tensors (constructed from momentum four-vectors as, $p^{\langle\mu_1}\cdots p^{\mu_n\rangle}$) that are orthogonal to each other as,
\begin{align}
    \intp p^{\langle\mu_1} \cdots p^{\mu_n\rangle} p_{\langle\nu_1} \cdots p_{\nu_\ell\rangle} F(\ep) = \frac{n! \delta_{n\ell}}{(2n +1)!!} \Delta^{\mu_1\cdots\mu_n}_{\nu_1\cdots\nu_n} \intp \left(p\cdot\Delta\cdot p\right)^n F(\ep),
\end{align}
where, $F(\ep)$ is some arbitrary function of particle energy, $\ep$, provided the integrals converge. We note that Eqs.~\eqref{n_q-def}-\eqref{P-def} contain equilibrium variables and are defined in terms of $\faz$ and $\afaz$, which are given by,
\begin{equation}
    \begin{aligned}
        %1
        \faz &=\frac{1}{\exp(\beta~p^{\mu}u_{\mu} - \xi_a) + r_a} = \frac{1}{\exp(\beta \ep - \xi_a) + r_a}, \\
        %2
        \afaz &= \frac{1}{\exp(\beta ~p^{\mu}u_{\mu} + \xi_a) + r_a}= \frac{1}{\exp(\beta \ep + \xi_a) + r_a}, \label{f0-def}
    \end{aligned}
\end{equation}
where $r_a = 0,+1,-1$ for particles following Maxwell-Boltzmann, Fermi-Dirac and Bose-Einstein statistics respectively, $\xi_a = \mu_a/T$ is the ratio between the total chemical potential\footnote{Note that, for the bosonic distribution (in this case, for the gluons) the chemical potential is vanishing i.e., $\xi_g = 0$. Consequently we can write, ${\sum_a}{\!\!'} \xi_a \left(\cdots\right) = \sum_a \xi_a \left(\cdots\right)$.} ($\mu_a = \sum_q q_a \mu_q$) of the particles of species $a$ and the medium temperature ($T\equiv 1/\beta$), with $\mu_q$ being the chemical potential related to the conserved charge densities, $n_q$. On the other hand, the out-of-equilibrium variables defined in Eqs.~\eqref{Pi-def}-\eqref{pi^mn-def} are expressed in terms of $\dfa$ and $\dafa$, which are to be obtained from the Boltzmann equation,
\begin{equation}
    \begin{aligned}
        \left(p\cdot \partial\right) f_a &= C[f_a, \Bar{f}_a], \\
        \left(p\cdot \partial\right) \Bar{f}_a &= \Bar{C}[f_a, \Bar{f}_a],
    \end{aligned}
    \label{boltzmanneq}
\end{equation}
where as in Ref.~\cite{Das:2021bkz}, for the collision kernels  $C, \Bar{C}$ we use the relaxation time approximation (RTA). Then writing, $\dfa = \phi_a \faz \tfaz$ and $\dafa = \Bar{\phi}_a \afaz \tafaz$ the Boltzmann equation reduces to,
\begin{equation}
    \begin{aligned}
        \left(p\cdot \partial\right) f_a &= - \left(\ep/\taur\right) \phi_a \faz \tfaz, \\
        \left(p\cdot \partial\right) \Bar{f}_a &= - \left(\ep/\taur\right) \Bar{\phi}_a \afaz \tafaz, \label{Beq2}
    \end{aligned}
\end{equation}
where $\taur$ is the single relaxation time for all species\footnote{In principle, the relaxation times should depend on the particular species $a$, but we have shown in Ref.~\cite{Bhadury:2020ngq} that unless we make the relaxation times momentum-dependent, the detailed balance condition and conservation laws require us to equate these relaxation times.}. Here we have defined, $\tfaz = 1 - r_a \faz$ and, $\tafaz = 1 - r_a \afaz$. We can solve for the correction terms $(\phi_a, \Bar{\phi}_a)$ iteratively in an order-by-order manner.

In the process of solving the Boltzmann equation, we will require the evolution of the chemical potentials $(\xi_a)$, the medium temperature ($T$) and fluid four-velocity ($u^{\mu}$), which are obtained from Eqs.~\eqref{heq1a}, \eqref{heq2} and \eqref{heq3} as,
\begin{subequations}
    \begin{align}
        %1
        \Dot{\beta} &= \beta_\theta\, \theta + \left(D_{20}\right)^{-1} \Big[\sum_a \left(J_{20}^{a,-}/J_{10}^{a,+}\right) \left(\partial\cdot n_a\right) - \Pi \theta + \pi^{\mu\nu} \sigma_{\mu\nu}\Big] \label{db}, \\
        %2
        \Dot{\xi}_a &= \xi^{(a)}_\theta\, \theta + \frac{1}{J_{10}^{a, +}} \left[\left(\frac{J_{20}^{a,-}}{D_{20}}\right) \Big\{\sum_{a'} \left(J_{20}^{a',-}/J_{10}^{a',+}\right) \left(\partial\cdot n_{a'}\right) - \Pi \theta + \pi^{\mu\nu} \sigma_{\mu\nu}\Big\} - \left(\partial\cdot n_a\right) \right] \label{dxi}, \\
        %3
        \Dot{u}_\mu &= - \frac{\left(\nabla^{\mu}\beta\right)}{\beta}
        + \sum_q \frac{n_q \left(\nabla_\mu \xi_q\right)}{\beta \left(\varepsilon + P\right)} + \frac{(\nabla_\mu \Pi)}{\left(\varepsilon + P\right)} - \frac{\Delta_{\mu\alpha} \partial_\beta \pi^{\alpha\beta}}{\left(\varepsilon + P\right)}, \label{nabla_m-beta}
    \end{align}
\end{subequations}
where, $\beta_\theta$, $\xi^{(a)}_\theta$ can be expressed as,
\begin{subequations}\label{b,x-def}
    \begin{align}
        %1
        \beta_\theta &= \left(D_{20}\right)^{-1} \Big[\sum_a \left(J_{20}^{a,-}/J_{10}^{a,+}\right) n_a - \left(\varepsilon + P\right)\Big], \label{beta_theta} \\
        %2
        \xi^{(a)}_\theta &= \left(\beta_\theta J_{20}^{a,-} - n_a\right)/J_{10}^{a,+}. \label{xi_theta}
    \end{align}
\end{subequations}
Here we have introduced the $J$-type thermodynamic integrals ($J_{nq}^{a,\pm}$), whose definition can be obtained from Eq.~\eqref{Inq} by replacing $(\faz\pm\afaz)$ with $(\faz \tfaz \pm \afaz \tafaz)$, i.e.,

\begin{align}
    J_{nq}^{a,\pm} \equiv \frac{1}{(2q+1)!!} \intp \ep^{n-2q} \left(p\cdot\Delta\cdot p\right)^q \left(\faz \tfaz \pm \afaz \tafaz\right), \label{Jnq}
\end{align}
which obey the following recursion relations:
\begin{subequations}
    \begin{align}
        %1
        J_{nq}^{a,\pm} &= \frac{1}{\beta} \Big[(n-2q) I_{n-1,q}^{a,\pm} - I_{n-1,q-1}^{a,\pm}\Big] \hspace{2cm} (a = u,d,s,g), \label{rec1} \\
        %2
        J_{nq}^{a,\pm} &= \frac{1}{(2 q + 1)} \Big[m^2 J_{n-2,q-1}^{a,\pm} - J_{n,q-1}^{a,\pm}\Big] \hspace{2cm} (a = u,d,s,g), \label{rec2}
    \end{align}
\end{subequations}
where we note the recursion relation similar Eq.~\eqref{rec2} is also satisfied by the thermodynamic integral, $I_{nq}^{a,\pm}$ as well as the ones introduced in Appendix \ref{app:TI}. In Eqs.~\eqref{db}, \eqref{dxi} and, \eqref{beta_theta} we encounter the term $D_{20}$, which can be obtained using the definition,
\begin{align}
    D_{nq} = \Big[\sum_a \left(J_{nq}^{a,-} J_{nq}^{a,-}/J_{n-1,q}^{a,+}\right) - J_{n+1,q}^{+}\Big]. \label{D_nq}
\end{align}

We are now ready to solve the Boltzmann equations \eqref{Beq2} with the help of Eqs.~\eqref{db}-\eqref{nabla_m-beta}. Once we have the expressions for $\phi_a, \Bar{\phi}_a$ we can determine the entropy production in the system. Thus, using the Boltzmann's H-theorem, the entropy current for particles following quantum statistics is given by \cite{DeGroot:1980dk, Jaiswal:2013fc},
\begin{align}
    H^\mu &=  - {\sum_a}' \intp p^\mu \Big[\left(f_a \ln f_a + r_a \Tilde{f}_a \ln\Tilde{f}_a\right) + \left(\Bar{f}_a \ln \Bar{f}_a + r_a \Tilde{\Bar{f}}_a \ln\Tilde{\Bar{f}}_a\right)\Big],
    \label{H^mu}
\end{align}
where $r_a = \pm 1$. The entropy production can be obtained by taking the divergence of the entropy current from Eq.~\eqref{H^mu} and using the Boltzmann equations Eq.~\eqref{boltzmanneq}, as
\begin{align}
    \partial_\mu H^\mu = - {\sum_a}' \intp \left[C_a [f_a, \Bar{f}_a] \ln\left(1 + \frac{\phi_a}{1 - r \phi_a \faz}\right) + \Bar{C}_a [f_a, \Bar{f}_a] \ln\left(1 + \frac{\Bar{\phi}_a}{1 - r \Bar{\phi}_a \afaz}\right)\right]. \label{d.H}
\end{align}
This will allow us to determine the dissipative currents at first-order and their second-order evolution equations along with the corresponding transport coefficients.

%~~~~~~~~~~~~~~~~~~~~~~~~~~~~~~~~~~~~~~~~~~~~
\subsection{First-order transport properties}
\label{ssec:FOTP}
%~~~~~~~~~~~~~~~~~~~~~~~~~~~~~~~~~~~~~~~~~~~~

The first order correction to the phase-space distribution functions, i.e., $\phi_{(1)}^a$, and $\Bar{\phi}_{(1)}^a$ can be obtained from the Boltzmann equation by using the Chapman-Enskog like iterative method~\cite{DeGroot:1980dk,Jaiswal:2013npa}. Therefore, keeping terms up to first order in spacetime gradients, we can obtain the off-equilibrium corrections to distribution functions from Eq.~\eqref{Beq2} as,
\begin{subequations}\label{phi1-def}
    \begin{align}
        %1
        \phi_{(1)}^a &= \taur \!\left[\!A_{a,\textbf{p}} \theta +\! \sum_{q} B^{aq}_{\textbf{p}} p^{\ab{\mu}} \left(\nabla_\mu\xi_q\right) + \beta \ep^{-1} p^{\langle\mu} p^{\nu\rangle} \sigma_{\mu\nu} \right], \label{phi_1^a-O1}\\
        %2
        \Bar{\phi}_{(1)}^a &= \taur \!\left[\!\Bar{A}_{a,\textbf{p}} \theta + \sum_{q} \Bar{B}^{aq}_{\textbf{p}} p^{\ab{\mu}} \left(\nabla_\mu\xi_q\right) + \beta \ep^{-1} p^{\langle\mu} p^{\nu\rangle} \sigma_{\mu\nu} \right]. \label{aphi_1^a-O1}
    \end{align}
\end{subequations}
where, the index $a$ indicates up ($u$), down ($d$), strange ($s$) quarks and gluons ($g$). However since gluons do not carry any $B, Q, S$ charges, hence for gluons, $B_{\textbf{p}}^{gq} = \Bar{B}_{\textbf{p}}^{gq} = 0$. In Eqs.~\eqref{phi1-def} we have used the following notations,
\begin{subequations}
    \begin{align}
        %1
        A_{a,\textbf{p}} &= \left(\beta_\theta - \frac{\beta}{3}\right) \ep + \frac{\beta m^2}{3 \ep} - \xi_\theta^{(a)},
        \hspace{2cm}
        \Bar{A}_{a,\textbf{p}} = \left(\beta_\theta - \frac{\beta}{3}\right) \ep + \frac{\beta m^2}{3 \ep} + \xi_\theta^{(a)}, \\
        %2
        B^{aq}_{\textbf{p}} &= \frac{n_{q}}{\left(\varepsilon + P\right)} - \frac{q_a}{\ep},
        \hspace{4.2cm}
        \Bar{B}^{aq}_{\textbf{p}} = \frac{n_{q}}{\left(\varepsilon + P\right)} + \frac{q_a}{\ep},
    \end{align}
\end{subequations}
with $\beta_\theta$ and, $\xi_\theta^{(a)}$ being already defined in Eqs.~\eqref{b,x-def}. Taking divergence of the entropy four current from Eq.~\eqref{H^mu} we can write the entropy production, keeping terms up to second order in spacetime gradient as \cite{Jaiswal:2013fc},
\begin{align}
    %1
    \partial_\mu H^\mu &= {\sum_a}' \intp \left(\ep/\taur\right) \left[\big(\phi^a_{(1)}\big)^2 \faz \tfaz + \big(\Bar{\phi}^a_{(1)}\big)^2 \afaz \tafaz \right] \\
    %2
    &= - \beta \Pi \theta - \sum_q n_q^\mu \left(\nabla_\mu \xi_q\right) + \beta \pi^{\mu\nu} \sigma_{\mu\nu}. \label{d.H2}
\end{align}
where,
\begin{align}
    \Pi = - \zeta \theta = - \taur \beta_\Pi \theta,
    \qquad\quad
    n_q^\mu = \sum_{q'} \kappa_{qq'} \left(\nabla^\mu\xi_{q'}\right) = \taur \sum_{q'} \beta_{qq'} \left(\nabla^\mu\xi_{q'}\right),
    \qquad\quad
    \pi^{\mu\nu} = 2 \eta \sigma^{\mu\nu} = 2 \taur \beta_{\pi} \sigma^{\mu\nu}, \label{NS-eqn}
\end{align}
are the relativistic Navier-Stokes equations. The coefficients, $\zeta, \kappa_{qq'}$ and, $\eta$ are bulk viscosity, charge conductivity\footnote{Note that, throughout the article we may use the terms `diffusion' and `conductivity' interchangeably.} (diffusion matrix) and shear viscosity respectively which given as,
\begin{subequations}
    \begin{align}
        %1
        \zeta &= \taur T {\sum_a}' \intp \ep \left[\left\{\left(\beta_\theta - \frac{\beta}{3}\right) \ep + \frac{\beta m^2}{3 \ep} - \xi^{(a)}_\theta\right\}^2 \faz \tfaz + \left\{ \left(\beta_\theta - \frac{\beta}{3}\right) \ep + \frac{\beta m^2}{3 \ep} + \xi^{(a)}_\theta\right\}^2\right] \afaz \tafaz, \label{zeta_QS} \\
        %2
        \kappa_{qq'} &=\! - \frac{\taur}{3} \!\sum_{a} \!\!\intp\! \frac{\left(p\cdot\Delta\cdot p\right)}{\ep} \!\left[\left\{q_{a} - \frac{n_{q} \ep}{\left(\varepsilon + P\right)}\right\} \!\left\{q'_{a} - \frac{n_{q'} \ep}{\left(\varepsilon + P\right)}\right\}\! \faz \tfaz - \left\{q_{a} + \frac{n_{q} \ep}{\left(\varepsilon + P\right)}\right\} \!\left\{q'_{a} + \frac{n_{q'} \ep}{\left(\varepsilon + P\right)}\right\}\! \afaz \tafaz\right], \label{kappa_qq'_QS} \\
        %3
        \eta &= \frac{\beta \taur}{15} {\sum_a}' \intp \frac{\left(p\cdot\Delta\cdot p\right)^2}{\ep} \left(\faz \tfaz +  \afaz \tafaz\right). \label{eta_QS}
    \end{align}
\end{subequations}
It is straightforward to note from Eqs.~\eqref{zeta_QS} and \eqref{eta_QS} that both $\zeta$ and $\eta$ are explicitly positive due to the squares in the integrands. Once again, from the above expressions, one can observe that gluons contribute to shear and bulk viscosity, but they do not contribute to the diffusion matrix elements ($\kappa_{qq^{\prime}}$). Similar observations can also be made for the diagonal components of $\kappa_{qq'}$, and we note that the expression of Eq.~\eqref{kappa_qq'_QS} coincides with the results from Refs.~\cite{Das:2021bkz, Dey:2024hhc}. Performing the momentum integrals in Eqs.~\eqref{zeta_QS}-\eqref{eta_QS} we can write the $\tau_R$ independent coefficients for the dissipative transport coefficients in Eq.~\eqref{NS-eqn} as
\begin{subequations}
    \begin{align}
        %1
        \beta_\Pi &=\! \frac{1}{\beta} {\sum_a}' \!\!\left[\!\left(\!\beta_\theta \!-\! \frac{\beta}{3}\!\right)^{\!2}\!\!\! J_{30}^{a,+} \!\!-\! 2 \xi^{(a)}_\theta \!\!\left(\!\beta_\theta \!-\! \frac{\beta}{3}\!\right)\! J_{20}^{a,-} \!\!\!+\! \left\{\!\!\left(\xi^{(a)}_\theta\!\right)^2 \!\!\!+\! \frac{2\beta m^2}{3} \!\left(\!\beta_\theta - \frac{\beta}{3}\right) \!\!\right\}\! J_{10}^{a,+} \!\!-\! \frac{2 \beta m^2 \xi^{(a)}_\theta\! J_{00}^{a,-}}{3} \!+\! \frac{\beta^2 m^4\! J_{-1,0}^{a,+}}{9}\right]\!, \label{beta_Pi} \\
        %2
        \beta_{qq'} &= - \left[\sum_{a} q_{a} q'_{a} J_{11}^{a,+} + \frac{T n_{q} n_{q'}}{\left(\varepsilon + P\right)}\right], \label{beta_qq'} \\
        %3
        \beta_\pi &= \beta J_{32}^{+}, \label{beta_pi}
    \end{align}
\end{subequations}
where, we have used Eqs.~\eqref{rec1} and, \eqref{rec2}. 
The expression for the diagonal components of $\beta_{qq'}$ from Eq.~\eqref{beta_qq'} shares resemblance with the expression of baryon charge conductivity of Ref.~\cite{Bhadury:2020ngq}.

%~~~~~~~~~~~~~~~~~~~~~~~~~~~~~~~~~~~~~~~~~~~~~
\subsection{Second-order transport properties}
\label{ssec:SOTP}
%~~~~~~~~~~~~~~~~~~~~~~~~~~~~~~~~~~~~~~~~~~~~~

To study the second order evolution equation of the dissipative currents, we use the results from Eq.~\eqref{phi1-def} and substitute into the Boltzmann equation, \eqref{Beq2}. Keeping terms up to second order in spacetime gradients, we obtain the second-order correction to the phase-space distribution function as,
\begin{subequations}\label{phi2-res}
    \begin{align}
        %1
        \phi_{(2)}^a &= \taur \!\left[ \sum_{a'} R^{aa'}_{\textbf{p}} \left(\partial\cdot n_{a'}\right) + S^{a}_{\textbf{p}} \Big\{(\pi:\sigma) -\Pi \theta\Big\} - \frac{\beta p_{\ab{\alpha}} \! \left(\partial_\beta \pi^{\alpha\beta} - \nabla^\alpha\Pi\right)}{ \left(\varepsilon+P\right)} \right] \nonumber\\
        %2
        &\hspace{2cm}- \frac{\left(\taur/\ep\right)}{\faz \tfaz} (p\cdot\partial) \left[\left\{\left(\frac{A_{a,\textbf{p}}}{\beta_\Pi}\right) \Pi + \sum_{q,q'} B^{aq}_{\textbf{p}}\, \hat{\beta}_{qq'} p_{\ab{\mu}} n_{q'}^\mu + \left(\frac{\beta}{2 \ep \beta_\pi}\right) p^{\langle\mu} p^{\nu\rangle} \pi_{\mu\nu}\right\} \faz \tfaz\right], \label{phi_1^a-O2}\\
        %3
        \Bar{\phi}_{(2)}^a &= \taur \!\left[\sum_{a'} \Bar{R}^{aa'}_{\textbf{p}} \left(\partial\cdot n_{a'}\right) + \Bar{S}^{a}_{\textbf{p}} \Big\{(\pi:\sigma) -\Pi \theta\Big\} - \frac{\beta p_{\ab{\alpha}} \! \left(\partial_\beta \pi^{\alpha\beta} - \nabla^\alpha\Pi\right)}{\left(\varepsilon+P\right)} \right] \nonumber\\
        %4
        &\hspace{2cm}- \frac{\left(\taur/\ep\right)}{\afaz \tafaz} (p\cdot\partial) \left[\left\{\left(\frac{\Bar{A}_{a,\textbf{p}}}{\beta_\Pi}\right) \Pi + \sum_{q,q'} \Bar{B}^{aq}_{\textbf{p}}\, \hat{\beta}_{qq'} p_{\ab{\mu}} n_{q'}^\mu + \left(\frac{\beta}{2 \ep \beta_\pi}\right) p^{\langle\mu} p^{\nu\rangle} \pi_{\mu\nu}\right\} \afaz \tafaz\right]. \label{aphi_1^a-O2}
    \end{align}
\end{subequations}
where,
\begin{subequations}
    \begin{align}
        %1
        R^{aa'}_{\textbf{p}} &= \left(\frac{J_{20}^{a',-}}{D_{20} J_{10}^{a',+} }\right) \left(\ep - \frac{J_{20}^{a,-}}{J_{10}^{a,+}} + \frac{D_{20} J_{10}^{a',+}}{J_{20}^{a',-} J_{10}^{a,+}}\right),
        \hspace{1cm}
        \Bar{R}^{aa'}_{\textbf{p}} = \left(\frac{J_{20}^{a',-}}{D_{20} J_{10}^{a',+} }\right) \left(\ep + \frac{J_{20}^{a,-}}{J_{10}^{a,+}} - \frac{D_{20} J_{10}^{a',+}}{J_{20}^{a',-} J_{10}^{a,+}}\right), \label{Raa'-def}\\
        %2
        S^{a}_{\textbf{p}} &= \left(D_{20}\right)^{-1}\! \left(\ep - \frac{J_{20}^{a,-}}{J_{10}^{a,+}} \right)
        \hspace{4.3cm}
        \Bar{S}^{a }_{\textbf{p}} = \left(D_{20}\right)^{-1}\! \left(\ep + \frac{J_{20}^{a,-}}{J_{10}^{a,+}} \right). \label{Sa-def}
    \end{align}
\end{subequations}

Starting from Eq.~\eqref{H^mu}, the entropy production can be given by \cite{Jaiswal:2013fc},
\begin{align}
    %1
    \partial_\mu H^\mu &=  \sum_a \intp \left(\ep/\taur\right) \left[\big(\phi^a_{(1)}\big)^2 \faz \tfaz + \big(\Bar{\phi}^a_{(1)}\big)^2 \afaz \tafaz\right] + 2 \sum_a \intp \left(\ep/\taur\right) \left(\phi^a_{(1)} \phi^a_{(2)} \faz \tfaz + \Bar{\phi}^a_{(1)} \Bar{\phi}^a_{(2)} \afaz \tafaz\right) \nonumber\\
    %2
    &\hspace{1.5cm}+ \sum_a \intp \left(\ep/\taur\right) \left[\big(\phi^a_{(1)}\big)^3 \faz \tfaz \left(\tfaz - \faz/2\right) + \big(\Bar{\phi}^a_{(1)}\big)^3 \afaz \tafaz \left(\tafaz - \faz/2\right)\right] + \mathcal{O} (\partial^4), \label{d.H3}
\end{align}
where we have kept terms up to third order in spacetime gradients. Substituting the results for $\phi_{(1)}^a, \phi_{(2)}^a$ and, $\Bar{\phi}_{(1)}^a, \Bar{\phi}_{(2)}^a$ from Eqs.~\eqref{phi1-def}, \eqref{phi2-res} we can find the entropy production to be given by,
\begin{align}
    %1
    \partial_\mu \mathcal{H}^\mu &= - \beta \Pi \!\bigg[\theta \!+\! \beta_0 \Dot{\Pi} \!+\! \beta_{\Pi\Pi} \Pi \theta \!+\! \beta_{\pi\pi} \pi^{\mu\nu} \sigma_{\mu\nu} \!+\! \sum_{q'} \psi^{q'}_{n\mathfrak{a}} n_{q'}^\mu \Dot{u}_\mu \!+\! \sum_{q,q''} \psi^{qq''}_{nn} n_{q''}^\mu \left(\nabla_\mu \xi_q\right) \!+\! \sum_{a'} \psi_{n}^{a'} \left(n_{a'}^\mu \Dot{u}_\mu \!-\! \nabla_\mu n_{a'}^\mu\right) \nonumber\\
    & ~~~~~~~~~~~~~~~~~~~~~~~~~~~~~~~~~~~~~~~~~~~~~~~~~~~~~~~~~~~+ \sum_{q'} \psi_{n}^{q'} \left(\nabla \cdot n_{q'}\right)\bigg] \nonumber\\
    %2
    - \beta &\!\sum_{q,q'}\!\! n_{q'}^\mu \!\!\bigg[T \! \left(\nabla_\mu \xi_{q}\right)\! \delta_{qq'} \!-\! \beta_1^{qq'} \!\! \Dot{n}_\mu^q \!-\! \beta_{n \Pi}^{qq'} n_\mu^q \theta \!+\! \beta_{\pi}^{qq'} \!\!\left(\nabla^\nu \pi_{\ab{\mu\nu}}\right) \!-\!\! \sum_{q''}\! \beta_{\pi n}^{qq'\!q''} \!\pi_{\mu\nu} \!\left(\nabla^\nu \xi_{q''}\right) \!-\!\! \sum_{q''}\! \beta_{\Pi n}^{qq'\!q''} \! \Pi \!\left(\nabla_\mu \xi_{q''}\right) \!\nonumber\\
    & ~~~~~~~~~~~~~~~~~~~~~~~~~~~~~~~~~~~~~~~~~~~~~~~~~~~+\! \psi_{\pi}^{qq'} \!\!\left(\nabla^\nu \pi_{\mu\nu}\right) \!+\! \psi_{\Pi}^{qq'} \!\left(\nabla_\mu \Pi\right) \!\bigg] \nonumber\\
    %3
    + \beta &\pi_{\mu\nu} \bigg[\sigma^{\mu\nu} \!-\! \beta_2 \Dot{\pi}^{\ab{\mu\nu}} \!-\! \beta_{\pi\Pi} \pi^{\mu\nu} \theta \!-\! \sum_{q,q''} \varphi_{nn}^{qq'}\, n_{q''}^{\langle\mu} \big(\nabla^{\nu\rangle} \xi_q\big) \!-\! \sum_{q'} \varphi_{n\mathfrak{a}}^{q'}\, n_{q'}^{\langle\mu} \Dot{u}^{\nu\rangle} \!+\! \varphi_{\pi\pi} \frac{14}{5} \pi^{\alpha\langle\mu} \sigma_{~\,\alpha}^{\nu\rangle} \!\nonumber\\
    & ~~~~~~~~~~~~~~~~~~~~~~~~~~~~~~~~~~~~~~~~~~~~~~+\! \alpha_{\pi\pi} \pi^{\nu\langle\mu} \sigma^{\alpha\rangle}_{~\,\alpha} \!+\! \sum_{q'} \alpha_{n}^{q'} \big(\nabla^{\langle\mu} n_{q'}^{\nu\rangle} \big) \bigg], \label{d.S-fin}
\end{align}
where the various coefficients appearing here are defined in Appendix \ref{app:VTC}. Demanding the positivity of entropy production, we can write,
\begin{align}
    %1
    \Pi &= - \zeta \left[\theta \!+\! \beta_0 \Dot{\Pi} \!+\! \beta_{\Pi\Pi} \Pi \theta \!+\! \beta_{\pi\pi} \pi^{\mu\nu} \sigma_{\mu\nu} \!+\! \sum_{q'} \psi^{q'}_{n\mathfrak{a}} n_{q'}^\mu \Dot{u}_\mu \!+\! \sum_{q,q''} \psi^{qq''}_{nn} n_{q''}^\mu \left(\nabla_\mu \xi_q\right)\right. \nonumber\\
    %2
    &\hspace{2cm}\left.+\! \sum_{a'} \psi_{n}^{a'} \left(n_{a'}^\mu \Dot{u}_\mu \!-\! \nabla_\mu n_{a'}^\mu\right) + \sum_{q'} \psi_{n}^{q'} \left(\nabla \cdot n_{q'}\right)\right], \label{Pi-O2}\\
    %3
    n_{q'}^\mu &= \sum_{q} \lambda_{qq'} \left[T \! \left(\nabla^\mu \xi_{q}\right) \!-\! \beta_1^{qq'} \!\! \Dot{n}^\mu_q \!-\! \beta_{n \Pi}^{qq'} n^\mu_q \theta \!+\! \beta_{\pi}^{qq'} \!\!\left(\nabla_\nu \pi^{\ab{\mu\nu}}\right) \!-\!\! \sum_{q''}\! \beta_{\pi n}^{qq'\!q''} \!\pi^{\mu\nu} \!\left(\nabla_\nu \xi_{q''}\right)\right. \nonumber\\
    %4
    &\hspace{2cm}\left.-\!\! \sum_{q''}\! \beta_{\Pi n}^{qq'\!q''} \! \Pi \!\left(\nabla^\mu \xi_{q''}\right) \!+\! \psi_{\pi}^{qq'} \!\!\left(\nabla_\mu \pi^{\mu\nu}\right) \!+\! \psi_{\Pi}^{qq'} \!\left(\nabla^\mu \Pi\right) \!\right], \label{nq-O2} \\
    %5
    \pi^{\mu\nu} &= 2 \eta \left[\sigma^{\mu\nu} \!-\! \beta_2 \Dot{\pi}^{\ab{\mu\nu}} \!-\! \beta_{\pi\Pi} \pi^{\mu\nu} \theta \!-\! \sum_{q,q''} \varphi_{nn}^{qq'}\, n_{q''}^{\langle\mu} \big(\nabla^{\nu\rangle} \xi_q\big) \!-\! \sum_{q'} \varphi_{n\mathfrak{a}}^{q'}\, n_{q'}^{\langle\mu} \Dot{u}^{\nu\rangle}\right. \nonumber\\
    %6
    &\hspace{2cm}\left.+\! \frac{14}{5} \varphi_{\pi\pi} \pi^{\alpha\langle\mu} \sigma_{~\,\alpha}^{\nu\rangle} \!+\! \alpha_{\pi\pi} \pi^{\nu\langle\mu} \sigma^{\alpha\rangle}_{~\,\alpha} \!+\! \sum_{q'} \alpha_{n}^{q'} \big(\nabla^{\langle\mu} n_{q'}^{\nu\rangle} \big) \right], \label{pi-O2}
\end{align}
where, the coefficients $\zeta$, $\eta$ and, $\lambda_{qq'}$ are proportionality factors. For these evolution equations to reduce to the Navier-Stokes limit when truncated at first order, we must identify, $\zeta = \taur \beta_\Pi, \lambda_{qq'} = \taur \beta_{qq'}/T$, and  $\eta = \taur \beta_\pi$.  Then the evolution equation of the dissipative currents are given by,
\begin{align}
    %1
    \tau_\Pi \Dot{\Pi} \!+\! \Pi &= - \zeta \theta - \delta_{\Pi\Pi} \Pi \theta \!+\! \lambda_{\Pi\pi} \pi^{\mu\nu} \sigma_{\mu\nu} \nonumber\\
    %2
    &\hspace{1.8cm}-\! \sum_{q'} \left[\ell_{\Pi n}^{(q')} \left(\nabla \cdot n_{q'}\right) \!+\! \tau^{(q')}_{\Pi n} n_{q'}^\mu \Dot{u}_\mu\right] -\! \sum_{q,q''} \lambda^{(q,q'')}_{\Pi n} n_{q''}^\mu \left(\nabla_\mu \xi_q\right) \!+\! \sum_{a'} \lambda_{\Pi n}^{(a')} \left(n_{a'}^\mu \Dot{u}_\mu \!-\! \nabla_\mu n_{a'}^\mu\right), \label{Pi-evol} \\
    %3
    \sum_{q}\! \tau_{qq'} \Dot{n}^\mu_q \!+\! n_{q'}^\mu &=\!\! \sum_{q}\! \left[\kappa_{qq'} \! \left(\nabla^\mu \xi_{q}\right) \!-\! \delta_{nn}^{(q,q')} n^\mu_q \theta \!+\! \beta_{\pi}^{qq'} \!\left(\nabla_\nu \pi^{\ab{\mu\nu}}\right) \right], \label{nq-evol} \nonumber\\
    %4
    &\hspace{1,8cm}-\! \sum_{q''}\! \left[\lambda_{n\Pi}^{(q',q'')} \Pi \left(\nabla^\mu \xi_{q''}\right) \!+\! \lambda_{n\pi}^{(q',q'')} \pi^{\mu\nu} \left(\nabla_\nu \xi_{q''}\right) \right] -\! \ell_{n\Pi}^{(q')} \!\left(\nabla^\mu \Pi\right) \!+\! \ell_{n\pi}^{(q')} \!\left(\nabla_\nu \pi^{\mu\nu}\right)\\
    %5
    \tau_{\pi} \Dot{\pi}^{\ab{\mu\nu}} \!+\! \pi^{\mu\nu} &=\! 2 \eta \sigma^{\mu\nu} -\! \delta_{\pi\pi} \pi^{\mu\nu} \theta \!-\! \tau_{\pi\pi} \pi^{\alpha\langle\mu} \sigma_{~\,\alpha}^{\nu\rangle} \nonumber\\
    %6
    &\hspace{1.8cm}- \sum_{q,q''}\! \lambda_{\pi n}^{(q,q')}\, n_{q''}^{\langle\mu} \big(\nabla^{\nu\rangle} \xi_q\big) -\! \sum_{q'} \left[\tau_{\pi n}^{(q')}\, n_{q'}^{\langle\mu} \Dot{u}^{\nu\rangle} - \ell_{\pi n}^{(q')} \big(\nabla^{\langle\mu} n_{q'}^{\nu\rangle} \big)\right] \!+\! \lambda_{\pi\pi} \pi^{\nu\langle\mu} \sigma^{\alpha\rangle}_{~\,\alpha}, \label{pi-evol}
\end{align}
where the coefficients are defined in Appendix \ref{app:VTC}. In the theory of hydrodynamics often one encounters acausality and instability~\cite{Hiscock:1983zz, Hiscock:1985zz, Hiscock:1987zz} which is cured by the relaxation type second or higher order hydrodynamic equations under linear as well as non-linear perturbations \cite{Israel:1979wp, Koide:2006ef, Baier:2007ix, El:2009vj, Peralta-Ramos:2009srp, Denicol:2012cn, Bemfica:2019knx, Miron-Granese:2020mbf, Panday:2024hqp, Ye:2024phs}. The second order hydrodynamic equations as given in Eqs.~\eqref{Pi-evol}-\eqref{pi-evol} contains more transport coefficients as compared to the Naiver-Stokes limit, i.e., Eq.~\eqref{NS-eqn}. These transport coefficients can crucially determine the casusal structure of the hydrodynamic framework which needs to be analyzed separately. Numerical estimation of these transport coefficients will also be important for the application of the hydrodynamic framework for phenomenological studies. Since one of the important goals of the present calculation is the estimation of the diffusion coefficients, in the next section we present the estimation of the first order diffusion matrix elements.

%-------------------------------
\section{Result and Discussions}
\label{sec:R&D}
%-------------------------------

\begin{figure}[t]
    \centering
    \includegraphics[width=1\linewidth]{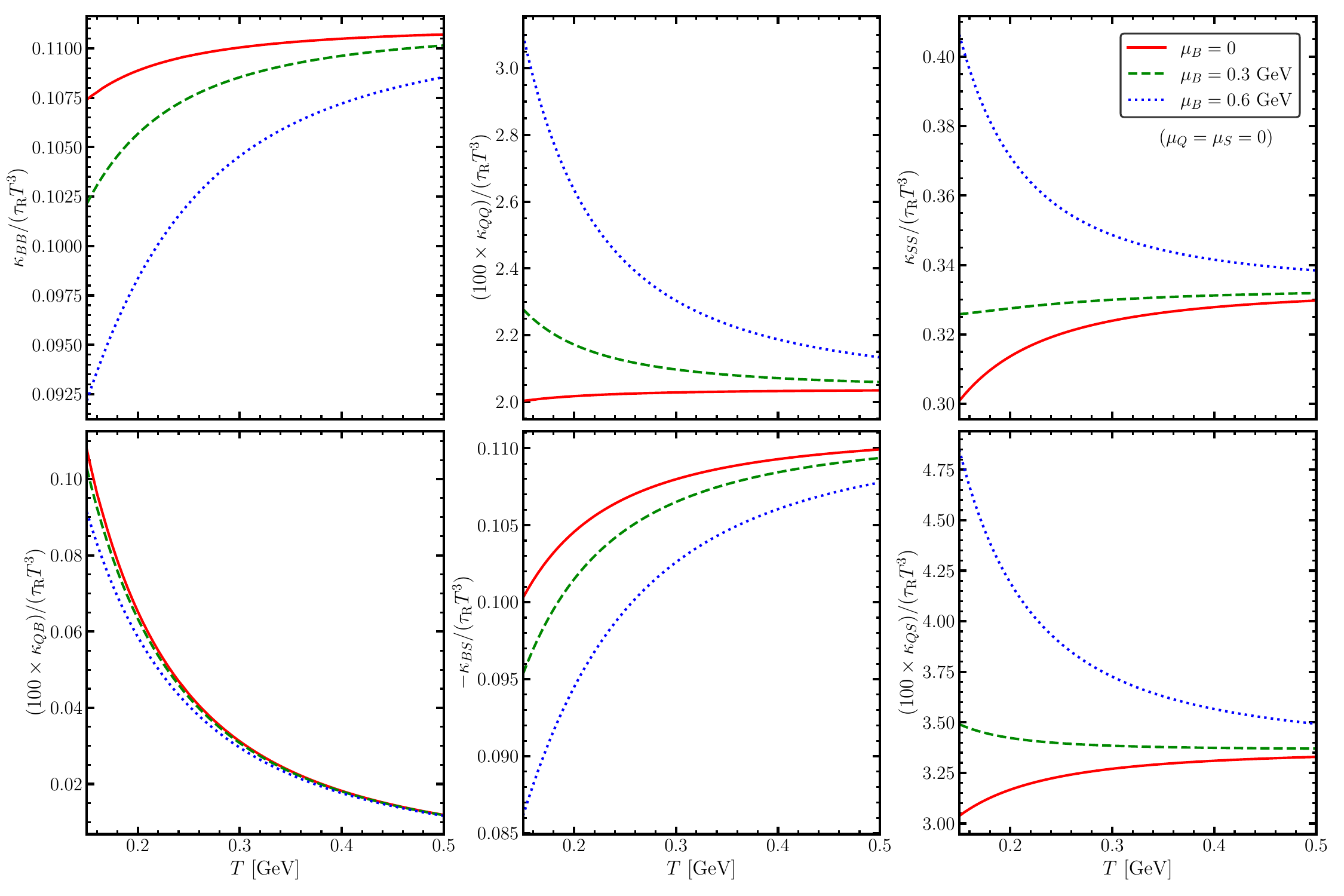}
    \caption{Temperature dependence of the scaled diagonal (first row) and off-diagonal (second row) components of the conductivity matrix (diffusion matrix), $\kappa_{qq'}/(\taur T^3)$ for $\mu_Q = \mu_S = 0$ values at $\mu_B = 0$ (solid lines), $\mu_B = 300$ MeV (dashed lines) and $\mu_B = 600$ MeV (dotted lines). The mass of fermions have been taken to be $(m_u, m_d, m_s) = (1, 1, 80)$ MeV. In these plots we consider temperature in the range $T\in [150-500]$ MeV. We consider the lowest value of temperature to be 150 MeV, assuming that the quark hadron transition temperature at zero baryon chemical potential is 150 MeV.} 
    \label{fig:kqq'1}
\end{figure}

\begin{figure}[t]
    \centering
    \includegraphics[width=1\linewidth]{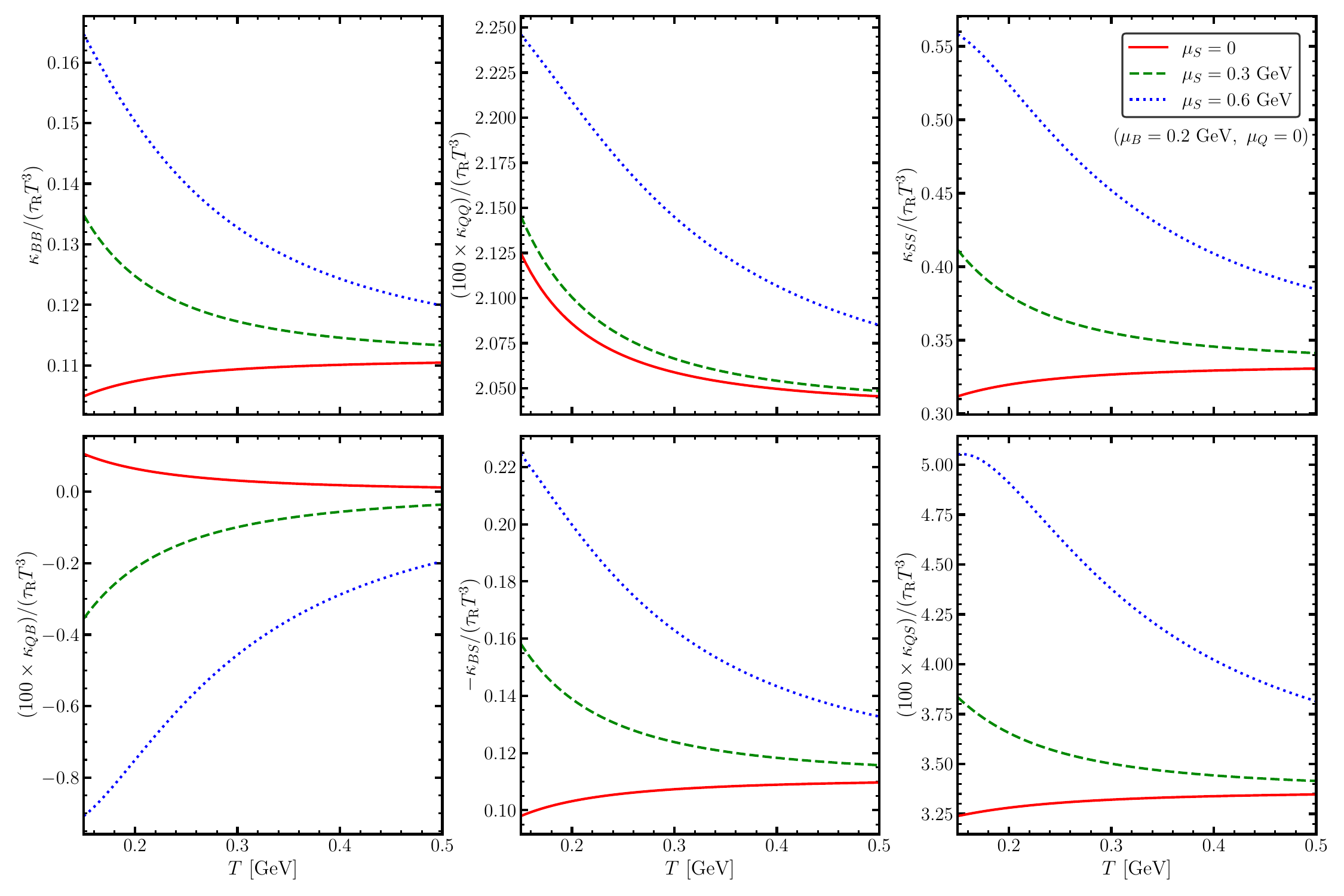}
    \caption{Temperature dependence of the scaled diagonal (first row) and off-diagonal components (second row) of the conductivity matrix (diffusion matrix), $\kappa_{qq'}/(\taur T^3)$ for $\mu_Q = 0$, $\mu_B = 300$ MeV values at $\mu_S = 0$ (solid lines), $\mu_S = 300$ MeV (dashed lines) and $\mu_S = 600$ MeV (dotted lines). The mass of fermions have been taken to be $(m_u, m_d, m_s) = (1, 1, 80)$ MeV.} 
    \label{fig:kqq'2}
\end{figure}

In Figs.~\ref{fig:kqq'1} and \ref{fig:kqq'2}, we depict the temperature dependence of the diagonal and off-diagonal components of the conductivity matrix in first and second rows respectively. From the expression of $\kappa_{qq'}$ in Eq.~\eqref{kappa_qq'_QS}, we note that the scaled conductivities ($\kappa_{qq'}/(\taur T^3)$), washes out the dependency on the relaxation time and the $T^3$, making the quantity dimensionless. Therefore, the crucial dependence on the temperature comes solely from the distribution function. For future references let us call the first term of Eq.~\eqref{beta_qq'} `the kinetic term', which contain $\faz (1 - \faz)$  (here, $a = u,d,s$) and the second term be dubbed as `the thermodynamic term', which is a function of thermodynamic variables only (such as, $n_q$, $\varepsilon$ and, $P$). Thus, the diffusion coefficient is a competition between these two terms. We also note that, Eq.~\eqref{beta_qq'} can be re-written as,
\begin{align}
    \beta_{qq'} = - \sum_{a,b} q_a q'_{b} \left[\delta_{ab} J_{11}^{a,+} + \frac{T n_a n_{b}}{(\varepsilon+P)}\right], \label{beta_qq'-2}
\end{align}
where, $a,b = u,d,s$ are the labels for particle species. Thus, apart from the competition between the kinetic term and the thermodynamic term, the properties of $\beta_{qq'}$ and hence $\kappa_{qq'}/(\taur T^3)$ also depend on the product of the charges i.e., $q_a q'_b$. In Fig.~\ref{fig:kqq'1} we keep the electric and strangeness chemical potentials fixed to zero values while varying the baryonic chemical potential for the values, $(0, 300, 600)$ MeV, which results in an increase of the particles of all three species ($u,d,s$) as compared to their anti-particles. We notice that in all the plots, at larger temperature values as $\mu_q/T \to 0$ the three curves show the tendency of merging together. This is due to the dominance of the thermal excitations.

In the upper left panel of Fig.~\ref{fig:kqq'1}, we plot $\kappa_{BB}/(\taur T^3)$, which increases with temperature for all values of $\mu_B$. This increasing behavior can be associated with temperature dependence of the distribution function, which can be understood in the Boltzmann limit.  Note that all the quarks have the same sign of baryonic charges ($B_u = B_d = B_s = + 1/3$). At zero baryonic chemical potential the thermodynamic term is zero. Whereas, for $\mu_B>0$, the thermodynamic term is finite and negative, as a result of which the conductivity is decreased. The variation of $\kappa_{BB}/(\taur T^3)$ is more evident in the low temperature range. 

In the upper central panel of Fig.~\ref{fig:kqq'1}, we have $\kappa_{QQ}/(\taur T^3)$. Since, the electric charges of the quarks follow, $Q_u = - 2 Q_d = - 2 Q_s = 2 e/3$ (where $e = \sqrt{4\pi/137}$), as $\mu_Q \to 0$, we find $n_Q \to 0$ and the thermodynamic term drops out as $\mu_S=0$, $\mu_B=0$. Thus, the behavior of $\kappa_{QQ}/(\taur T^3)$ is controlled by the kinetic term alone. At, $\mu_B=0$, we have $\faz = \afaz$ and hence the the conductivity has a finite value solely due to thermal excitations. As temperature is increased, the availability of charge carriers increase, causing the slow rise of $\kappa_{QQ}/(\taur T^3)$. At $\mu_B>0$, however, the density of quarks is more than anti-quarks and this dominance of quarks only increases with $\mu_B$. Consequently, the kinetic term attains a larger value. Here also in the high temperature range different curve merge, and the curves of $\kappa_{QQ}/(\taur T^3)$ for $\mu_B>0$ fall steadily to some constant value.

In the upper right panel of Fig.~\ref{fig:kqq'1}, we have $\kappa_{SS}/(\taur T^3)$, which is due to only the strange quarks as we have, $\mu_S = 0$ and $S_u = S_d = 0$. The qualitative features of the figure is same as the figure for $\kappa_{QQ}/(\taur T^3)$. Since only the strange quarks carry the strangeness quantum number, we find a simplified expression of $\kappa_{SS}$ leads to,
\begin{align}
    \kappa_{SS}/(\taur T^3) = - \frac{S_s^2}{T^3} \left[J_{11}^{s,+} + \frac{T n_s^2}{(\varepsilon + P)}\right]. \label{kss}
\end{align}
At $\mu_B = 0$, we have, $n_s = 0 \implies n_S = 0$ and hence only the kinetic term has a finite contribution. From Eq.~\eqref{kss} we observe that, while the kinetic term increases with $\mu_B$, the thermodynamic term, which suppresses $\kappa_{SS}$, does not grow as fast due to the enthalpy ($h = \varepsilon + P$) term in the denominator, which has contributions from all particles in the system and grows rather rapidly with increasing $\mu_B$. Thus, the kinetic term dominates causing the ratio, $\kappa_{SS}/(\taur T^3)$ to increases with $\mu_B$. As in the other two diagonal components, at large temperatures the lines for $\mu_B = (0, 300, 600)$ MeV merge together due to the dominance of thermal effects.

The left panel in the second row of Fig.~\ref{fig:kqq'1}, shows the temperature dependence of cross-diffusion coefficient (off diagonal component of diffusion matrix) $\kappa_{QB}/(\taur T^3)$ for three values of $\mu_B$ at zero electric and strangeness chemical potentials. The kinetic term is proportional to $(2 J_{11}^{u,+} -  J_{11}^{d,+} - J_{11}^{s,+})$. Consequently, the kinetic term suffers from a cancellation by $d$ and $s$-quarks, suppressing its values. The kinetic term increases slightly with $\mu_B$. At $\mu_B = 0$, the thermodynamic term drops out and, hence the $\kappa_{QB}/(\taur T^3)$ attains highest values at $\mu_B = 0$. On the other hand, at $\mu_B>0$, the finite value of the thermodynamic terms suppresses the cross-conductivity. The dependence of the thermodynamic terms on $\mu_B$ is relatively stronger as compared to the kinetic terms. Hence, even though the kinetic terms show slight increase with $\mu_B$, the thermodynamic terms suppresses the conductivities resulting in an overall suppression of $\kappa_{QB}/(\taur T^3)$ with $\mu_B$.

The central panel of the lower row of Fig.~\ref{fig:kqq'1} depicts, $\kappa_{BS}/(\taur T^3)$, which is negative as the kinetic term has contribution from $s$-quarks only, whose baryonic and strangeness charges have opposite signs. At $\mu_B = 0$, the thermodynamic term drops out, but at $\mu_B>0$ the thermodynamic term positively contributes to $\kappa_{BS}$ resulting in enhancement of the ratio, $\kappa_{BS}/(\taur T^3)$ and makes it less negative.

Finally, in the lower right panel of Fig.~\ref{fig:kqq'1}, we have shown the temperature dependence of $\kappa_{QS}/(\taur T^3)$. Qualitatively, the plot resembles the plots for $\kappa_{QQ}/(\taur T^3)$ and $\kappa_{SS}/(\taur T^3)$. Similarly to $\kappa_{BS}/(\taur T^3)$, the contribution to kinetic term comes only from the strange sector. Since both the electric and strangeness charges of the $s$-quarks are negative, their product leads to a positive contribution to $\kappa_{QS}/(\taur T^3)$. While, at $\mu_B = 0$, the thermodynamic term is zero, at $\mu_B>0$ the enthalpy in the denominator of the thermodynamic term grows rapidly, resulting in a suppression of the thermodynamic term. Consequently, the kinetic term play a dominant role, which increases with $\mu_B$.

In Fig.~\ref{fig:kqq'2}, we plot the same components of diffusion matrix as in Fig.~\ref{fig:kqq'1}, against temperature, while varying $\mu_S$ for the values, $(0, 300, 600)$ MeV at fixed $\mu_B = 200$ MeV and vanishing electric chemical potential ($\mu_Q = 0$). Unlike Fig.~\ref{fig:kqq'1}, where the baryonic chemical potential was varied causing the increase of densities of all the quarks, here we vary the strangeness chemical potential. Here, the kinetic term increases more rapidly as compared to the thermodynamic term with increasing $\mu_S/T$. This is due to the increased population of $s$-quarks, which affects the kinetic term more.

In the central panel of the first row in Fig.~\ref{fig:kqq'2}, we plot the variation of $\kappa_{QQ}/(\taur T^3)$ with temperature. Note that $\kappa_{QQ}$ and electrical conductivity are related. The scaled $\kappa_{QQ}$ or electric conductivity shows increment with $\mu_S$, which similarly to Fig.~\ref{fig:kqq'1}. The increase in $\mu_S$ directly boosts the population of $s$-quarks. Consequently, both the terms are equally enhanced at finite $\mu_S$ in comparison to their $\mu_S = 0$ values. However, since the kinetic term dominates the thermodynamic term, we notice an increase in the overall electrical conductivity with $\mu_S$.

In the upper right panel of Fig.~\ref{fig:kqq'2}, we have shown the temperature dependence of $\kappa_{SS}/(\taur T^3)$. The increased conductivity of strangeness charge with increasing $\mu_S$ at fixed $\mu_B$ and $\mu_Q$ is quite intuitive and follows from natural understanding of increased availability of strangeness carriers.

In the lower three plots of Fig.~\ref{fig:kqq'2}, we notice qualitative differences for $\kappa_{QB}/(\tau_RT^3)$ and $\kappa_{BS}/(\tau_RT^3)$, whereas $\kappa_{QS}/(\tau_RT^3)$ remains qualitatively same as $\kappa_{QS}/(\tau_RT^3)$ from Fig.~\ref{fig:kqq'1}. These features can be understood by noting that in Fig.~\ref{fig:kqq'2} as $\mu_S$ is increased, only the population of $s$-quarks is increased whose baryonic charge has opposite sign than its electrical and strangeness charges i.e., ${\rm sign} (Q_s) = {\rm sign} (S_s) = -$ but, ${\rm sign} (B_s) = +$. Therefore, we notice, at $\mu_S=0$, in the lower left and lower middle panels, the qualitative nature of the plots in Fig.~\ref{fig:kqq'2} is similar to those in Fig.~\ref{fig:kqq'1}. However, at $\mu_S>0$, as $s$-quarks start to dominate, both the scaled cross-conductivities, ($\kappa_{QB}/(\tau_RT^3)$ and $\kappa_{BS}/(\tau_RT^3)$) become more and more negative, resulting in a flip as compared to Fig.~\ref{fig:kqq'1}. The qualitative similarity in $\kappa_{QS}/(\tau_RT^3)$ can be understood from the fact that the electrical and strangeness charges of $s$-quarks are both of same sign. Thus, an increase in the population of $s$-quarks through increased $\mu_S$ is expected to boost the scaled cross-conductivity of $\kappa_{QS}/(\tau_RT^3)$.

\begin{figure}[t]
    \centering
    \includegraphics[width=\linewidth]{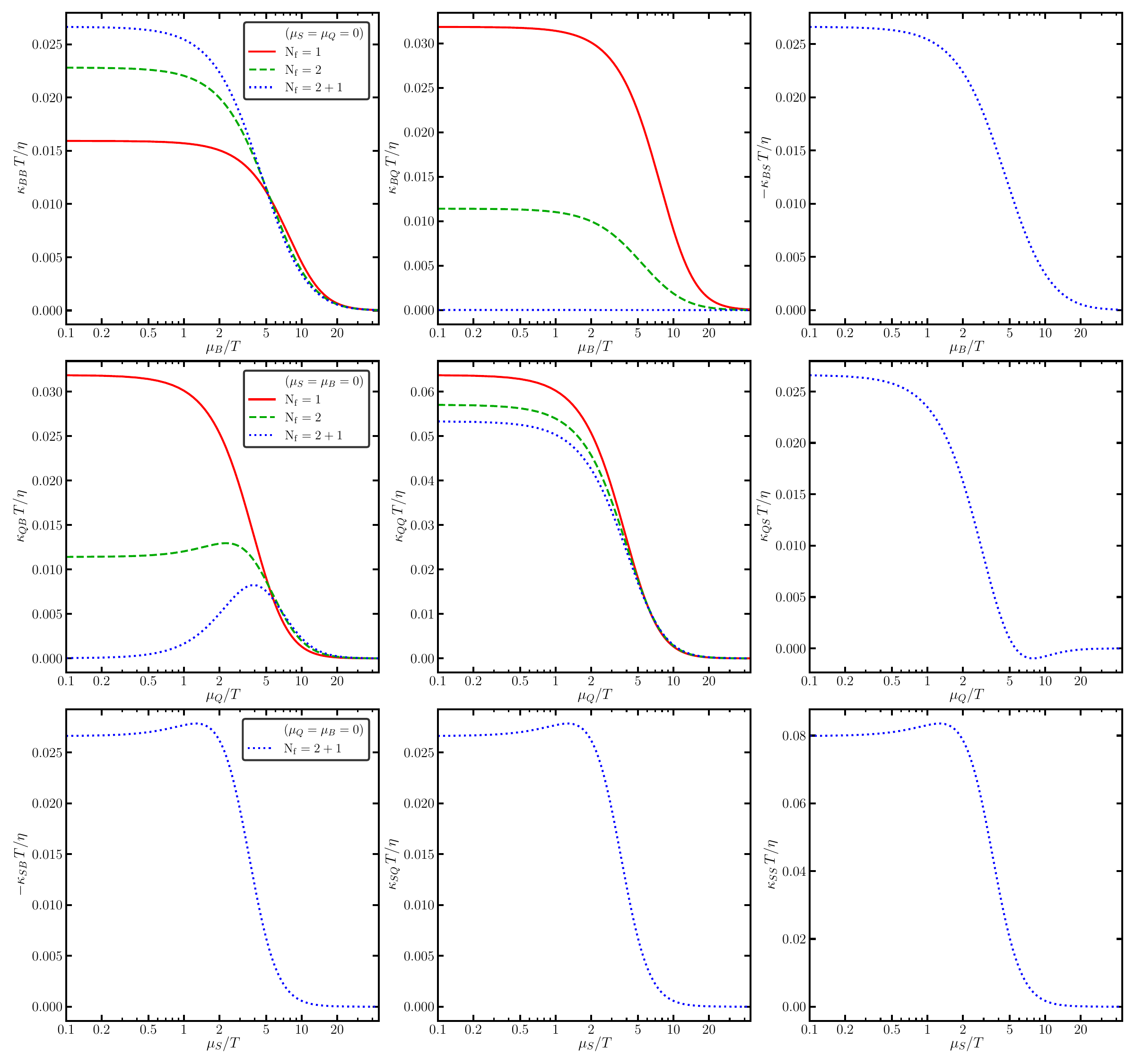}
    \caption{Ratios of charge conductivities to shear viscosity scaled by the temperature, $\kappa_{qq'} T/\eta$, for single (solid line), two (dashed line) and, $2+1$ (dotted lines) flavor massive quarks, plotted against $\mu_q/T$. The scaled mass of fermions have been taken to be $(z_u, z_d, z_s) = (0.001, 0.001, 0.08)$.}
    \label{fig:kqqTbe}
\end{figure}

In Fig.~\ref{fig:kqqTbe} we study the dimensionless ratio, $\kappa_{qq'} T/\eta$, which measures the importance of the various charge conductivities relative to the shear viscosity \cite{Jaiswal:2015mxa, Dey:2024hhc}, as functions of the ratios, $\xi_q \equiv \mu_q/T$. We observe that for all the cases, in the $\xi_q\to0$ limit, the ratios $\kappa_{qq'} T/\eta$ saturate to some constant values. On the other hand, as $\mu_q/T$ takes large values, the ratios, $\kappa_{qq'} T/\eta$ approaches a zero value. The drop to near zero values of the ratios, $\kappa_{qq'} T/\eta$ indicate the decreasing importance of the conductivities relative to the shear viscosity as the chemical potentials increase or equivalently, the temperature of the system decrease. The fall to zero can be understood from a set of factors. Although from Eq.~\eqref{f0-def}, it may appear that the increasing value of $\xi_a$ (through increasing $\xi_q$), should increase $\faz$, which in turn should increase the diffusion. However, there are other factors that causes the decrement. Firstly, in the second term of Eq.~\eqref{beta_qq'}, while $n_q$ increases with $\xi_q$, the factor in the denominator $(\varepsilon + P)$ increases much more rapidly, causing a suppression of the ratio $\kappa_{qq'} T/\eta$. Secondly, the increment in shear viscosity causes additional suppression. Lastly, Pauli-blocking also plays a major role in the suppression through the factor, $\faz (1-\faz)$, which approaches zero as $\xi_q$ takes larger value. The combination of these effects ensure the fall of $\kappa_{qq'} T/\eta$ with respect to $\xi_q$.

While in the cases of a single flavor, the drop to the zero value of the ratios $\kappa_{BB} T/\eta, \kappa_{BQ} T/\eta$ and, $\kappa_{QQ} T/\eta$ is monotonic, non-trivial features emerge in the cases of multiple flavor systems ($N_f = 2$ and, $2+1$). Since $u$ and $d$-quarks do not contain any strangeness charge, it is natural that $\kappa_{BS} T/\eta$, $\kappa_{QS} T/\eta$ and, $\kappa_{SS} T/\eta$ take zero values for $N_{\rm f} = 1, 2$ cases. Thus, their finite values and their non-trivial features for the $N_{\rm f} = 2 + 1$ case can be attributed to $s$-quark only, although the enthalpy ($h = \varepsilon + P$) and, shear viscosity ($\eta$) get contribution from all particle species.
The behavior of $\kappa_{BB} T/\eta$ is similar to Fig.~1 of Ref.~\cite{Jaiswal:2015mxa} and, Fig.~3 of Ref.~\cite{Das:2021bkz} i.e. they all decrease with increasing values of the chemical potential to temperature ratio and similar to Ref.~\cite{Jaiswal:2015mxa} the numerical values increase as the number of flavors is increased. This increase with addition of flavors is due to the fact that all three particles under consideration ($u,d,s$), have been assumed to carry baryonic charges of the same sign. However, these particles have different signs in their electric charges, i.e., $Q_u = +2e/3$ whereas, $Q_d = Q_s = -e/3$, ($e = \sqrt{4\pi/137}$). This negative sign appears in the exponential of the distribution functions of the thermodynamic integrals leading to a suppressed contribution to the electric conductivity. Consequently, the ratio $\kappa_{QQ} T/\eta$ is suppressed with increasing number of flavors as shown in Fig.~\ref{fig:kqqTbe}.

Compared to the ratio, $\kappa_{BB} T/\eta$, the flavor-dependence of the off-diagonal components $\kappa_{BQ} T/\eta=\kappa_{QB} T/\eta$ show an opposite behavior with increasing flavors, more in line with $\kappa_{QQ} T/\eta$. The suppression with increasing number of flavors may be attributed to the opposite signs of baryonic and electric charges in $d$ and, $s$-quarks as compared to $u$-quarks, which appear explicitly in the expression of $\beta_{qq'}$ as shown in Eq.~\eqref{beta_qq'-2}. In the case of $\kappa_{BQ}T/\eta$ against $\mu_B/T$ plot we observe a near zero value for $N_f = 2+1$ case. This can be understood by noting that at $\mu_Q = \mu_S = 0$ the kinetic and thermodynamic terms are proportional to $(2J_{11}^{u,+} - J_{11}^{d,+} - J_{11}^{s,+})$ and $(2n_u - n_d - n_s)$ respectively. Both of these terms nearly cancel each other.
It may be noted that, in $\kappa_{QB} T/\eta$ the curves for $N_f = 2$ and, $2+1$ cases, there are some non-monotonic behavior. This is a result of the competition between the kinetic and thermodynamic terms as well the fact that the $N_f = 2$ and, $2+1$ cases include $d$ and $d,s$-quarks respectively, that carry opposite electric and baryonic charges. 

\begin{figure}[t]
    \centering
    \includegraphics[width=0.55\linewidth]{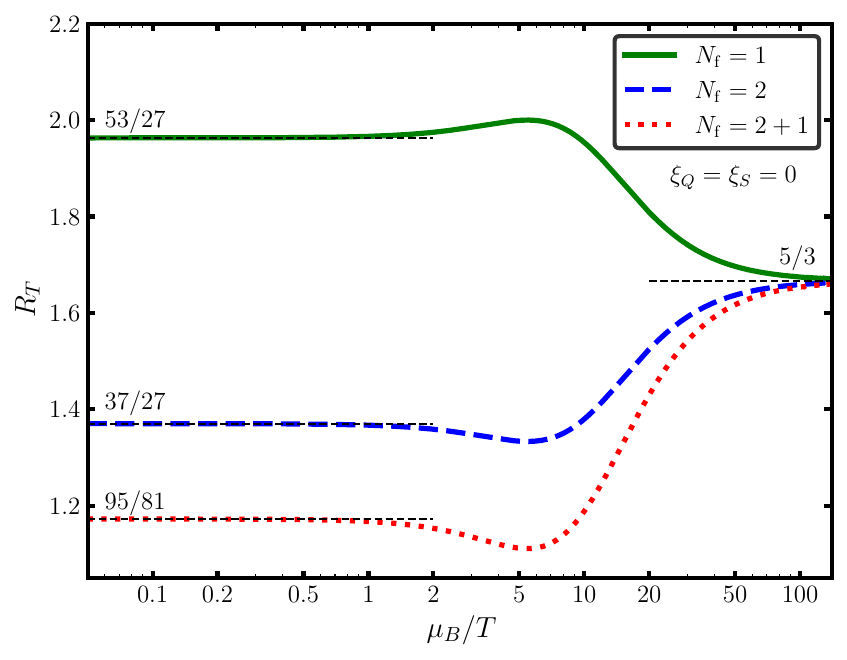}
    \caption{Ratios of thermal conductivity to shear viscosity, $\kappa_T/\eta$, scaled by the factor $\sum_a\mu_a^2/(\pi^2 T)$, for single (solid line), two (dashed line) and, $2+1$ (dotted lines) flavor massive quarks, plotted against $\mu_q/T$. The scaled mass of fermions have been taken to be $(z_u, z_d, z_s) = (0.001, 0.001, 0.08)$.}
    \label{fig:ktbe}
\end{figure}

In Fig.~\ref{fig:kqqTbe}, for the plots involving the strangeness such as $\kappa_{BS} T/\eta$, $\kappa_{QS} T/\eta$ and, $\kappa_{SS} T/\eta$ we obtain zero for the $N_f = 1, 2$ cases as they do not carry any strangeness charges. Hence we have plots only for the $N_f = 2+1$ cases. Note that these are not the same as the case with only $s$-quarks because of the enthalpy term in the denominator of Eq.~\eqref{beta_qq'}, which have contribution from all particle species including the gluons. The non-monotonic behavior of these plots are the result of the competition between the kinetic and thermodynamic terms.

Lastly, we study another dimensionless ratio \cite{Dey:2024hhc},
\begin{align}
    R_T = \frac{(\kappa_T/\eta)}{\pi^2 T} \sum_a \mu_a^2,
\end{align}
where
\begin{align}
    \kappa_T = \left(\!\frac{\varepsilon + P}{T \sum_q \mu_q n_q}\!\right)^{\!\!2} \!\! \sum_{q,q'} \mu_q \mu_{q'} \kappa_{qq'}, \label{kT}
\end{align}
is the heat flow for individual charge carriers. It is not to be confused with the heat flow coefficient related to $h^\mu = u_\beta \Delta^{\mu}_{\alpha} T^{\alpha\beta}$, which in the present case is zero due to our choice of Landau frame definition of fluid four-velocity. Similarly to Ref.~\cite{Dey:2024hhc} (which is a generalization to multiple conserved charge of Ref.~\cite{Jaiswal:2015mxa}) we also find scaling properties at small and large $\mu_q/T$. In Fig.~\ref{fig:ktbe} we plot $R_T$ against $\mu_B/T$. The values of $R_T$ at small $\mu_B/T$ are consistent with the AdS/CFT result \cite{Jain:2009pw}, $R_T = C_f$, where \cite{Jaiswal:2015mxa} $C_f = (4 g_g + 7 N_f g_f)/(9 N_f g_f)$, with $g_g$ and $g_f$ being the degeneracy of gluons and quarks, $N_f$ is the number of flavors considered. Hence, we get, $C_f = 53/27, 37/27$ and, $95/81$ for $N_f = 1, 2$ and, $2+1$ flavors respectively. At large $\mu_B/T$ we also find a flavor ($N_f$) independent value, $5/3$ similar to Ref.~\cite{Jaiswal:2015mxa, Dey:2024hhc}. 

%--------------------------------
\section{Conclusions and Outlook}
\label{sec:C&O}
%--------------------------------

In this article, we have derived the first-order as well as second-order hydrodynamic evolution equations for the dissipative currents for a system of multiple species - three massive fermions with multiple conserved charges (baryonic, electric and, strangeness) and massless, chargeless gluons using the entropy production in the kinetic theory framework. In particular we focused on the components of the diffusion matrix and examined its dependence on temperature and the chemical potentials through a dimensionless quantity, $\kappa_{qq'}/(\taur T^3)$. We also study various scaling properties of the diffusion coefficients as functions of the temperature scaled chemical potentials, $\mu_q/T$. By examining the relative importance of the diffusion coefficients with respect to the shear viscosity, we note that in the limit of large $\mu_q/T$, the shear viscosity dominates the transport properties over all the components of the diffusion matrix. While studying the properties of the scaled heat flow per charge carrier as a function of $\mu_B/T$, we find agreement with Ref.~\cite{Jaiswal:2015mxa} in both the large and small limits of $\mu_B/T$. In the intermediate range, the effect of mass and multiple conserved charges come into play resulting deviation from the single charge result of Ref.~\cite{Jaiswal:2015mxa}.

While the present work has been aimed at the system of QGP, it will be interesting to study the hadronic medium formed after the hadronization in heavy-ion collision experiments. We keep this study as a future work, where we shall perform some phenomenological investigations. While we have considered the conservation of particle species in the present work, which is appropriate for a system dominated by strong interaction, while studying the hadronic medium, we can relax this constraint and consider a system where only the charges are conserved. An important effect to study is the influence of magnetic field on various observables in the presence of multiple conserved charges. We may also investigate the consequence of multiple conserved charges for a spin polarizable medium as such systems has initiated renewed interest in the low collision energy regime, where $BQS$ physics may play significant role. While in the present work, we have limited ourselves to systems with momentum/energy-independent relaxation times, it will be useful to re-evaluate the evolution equations of the dissipative currents following Refs.~\cite{Rocha:2021zcw, Dash:2021ibx}. Such a theory will further help us in generalizing to a system of species-dependent relaxation times. In future we would also like to study the effect of realistic equation of state on the multi-charged system.

%-------------------------
\section*{Acknowledgments}
\label{sec:ack}
%-------------------------

S.B. would like to thank Dr. Amaresh Jaiswal, Dr. Sourav Dey and, Dr. Arghya Mukherjee for useful discussions. S. B. acknowledges the support from Anusandhan National Research Foundation (ANRF), India through National Post Doctoral Fellowship, File No. PDF/2025/004233. A.D. acknowledges the New Faculty Seed Grant (NFSG), NFSG/PIL/2024/P3825, provided by the Birla Institute of Technology and Science Pilani, Pilani Campus, India. A.D. acknowledges the Anusandhan National Research Foundation (ANRF), Advanced Research Grant (ARG), project number: ANRF/ARG/2025/000691/PS.

\appendix

%--------------------------------
\section{Thermodynamic Integrals}
\label{app:TI}
%--------------------------------

As we will see in Appendix-\ref{app:VTC}, to determine the second-order transport coefficients, we will have to define two new thermodynamic integrals, not defined in the main text, given by,
\begin{align}
    & K_{nq}^{a,\pm} = \frac{1}{(2q+1)!!} \intp \ep^{n-2q} \left(p\cdot\Delta\cdot p\right)^q \left(\faz \tfaz \hat{f}_a^{0} \pm \bar{f}_a^0 \tilde{\bar{f}}_a^0 \hat{\bar{f}}_a^0 \right), \label{K_nq}\\
    %4
    & L_{nq}^{a,\pm} = \frac{1}{(2q+1)!!} \intp \ep^{n-2q} \left(p\cdot\Delta\cdot p\right)^q \left(\faz \tfaz \tfaz \pm \afaz \tafaz \tafaz\right), \label{L_nq}
\end{align}    
where, $\hat{f}_a^{0} \equiv (1 - 2 r_a f_a^{0})$ and, $\hat{\Bar{f}}_a^{0} \equiv (1 - 2 r_a \Bar{f}_a^{0})$. We some expressions thermodynamic integrals in the following, that will be useful for analytical calculations for the case of massive particles in the limit of small chemical potentials,
\begin{align}
    %1
    I_{10}^{a,-} &= \left(\frac{g_a z_a^3 T^3}{4 \pi^2}\right) \sum_{j=1}^{\infty} \left(-r_a\right)^{j-1} \sinh\left(j \xi_a\right) \!\Big[K_3 (jz_a) - K_1 (jz_a) \Big], \label{I_10-expr}\\
    %2
    I_{20}^{a,+} &= \left(\frac{g_a z_a^4 T^4}{8 \pi^2}\right) \sum_{j=1}^{\infty} \left(-r_a\right)^{j-1} \cosh\left(j \xi_a\right) \!\Big[K_4 (jz_a) - K_0 (jz_a) \Big], \label{I_20-expr}\\
    %3
    I_{21}^{a,+} &= - \left(\frac{g_a z_a^4 T^4}{24 \pi^2}\right) \sum_{j=1}^{\infty} \left(-r_a\right)^{j-1} \cosh\left(j \xi_a\right) \!\Big[K_4 (jz_a) - 4 K_2 (jz_a) +3 K_0 (jz_a)\Big], \label{I_21-expr}\\
    %4
    J_{11}^{a,+} &= - \left(\frac{g_a z_a^3 T^3}{12 \pi^2}\right) \sum_{j=1}^{\infty} j \left(-r_a\right)^{j-1} \cosh\left(j \xi_a\right) \!\Big[K_3 (jz_a) - 5 K_1 (jz_a) + 4 K_{i,1} (jz_a)\Big], \label{J_11-expr}\\
    %5
    J_{32}^{a,+} &= \left(\frac{g_a z_a^5 T^5}{480 \pi^2}\right) \sum_{j=1}^{\infty} j \left(-r_a\right)^{j-1} \cosh\left(j \xi_a\right) \!\Big[K_5 (jz_a) - 7 K_3 (jz_a) + 22 K_1 (jz_a) - 16 K_{i,1} (jz_a)\Big], \label{J_32-expr}
\end{align}
where, $K_n(x)$ is the modified Bessel function of second kind and, $K_{i,1} (x) = $ is the first-order Bickley-Naylor function which can be expressed in terms of $K_n (x)$ and the Struve functions, $L_n(x)$ as,
\begin{align}
    K_{i,1} (jz_a) = \frac{\pi}{2} \Big[1 - z K_0 (j z_a) L_{-1} (j z_a) - z K_1 (j z_a) L_{0} (j z_a)\Big].
\end{align}

To derive the expressions in Eqs.~\eqref{I_10-expr}-\eqref{J_32-expr} we had to use the following definitions,
\begin{align}
    %1
    K_n (j z_a) &= \int_0^\infty d\theta \cosh n\theta\, \exp\left(- j z_a \cosh \theta\right), \\
    %2
    K_{i,n} (j z_a) &= \int_0^\infty d\theta \left(\sech \theta\right)^n \exp\left(- j z_a \cosh \theta\right).
\end{align}

In case of massless particles under the limit of small $\xi_a$, the thermodynamic integrals are given by,
\begin{align}
    %1
    I_{nq}^{a,+} &= \frac{g (-1)^q T^{n+2}}{(2q+1)!!\, \pi^2} \sum_{j=1}^\infty \frac{(-r)^{j-1}}{j^{n+2}} \cosh (j \xi_a) \Gamma(n+2), \\
    %2
    I_{nq}^{a,-} &= \frac{g (-1)^q T^{n+2}}{(2q+1)!!\, \pi^2} \sum_{j=1}^\infty \frac{(-r)^{j-1}}{j^{n+2}} \sinh (j \xi_a) \Gamma(n+2), \\
    %3
    J_{nq}^{a,+} &= \frac{g (-1)^q T^{n+2}}{(2q+1)!!\, \pi^2} \sum_{j=1}^\infty \frac{(-r)^{j-1}}{j^{n+1}} \cosh (j \xi_a) \Gamma(n+2), \\
    %4
    J_{nq}^{a,-} &= \frac{g (-1)^q T^{n+2}}{(2q+1)!!\, \pi^2} \sum_{j=1}^\infty \frac{(-r)^{j-1}}{j^{n+1}} \sinh (j \xi_a) \Gamma(n+2),
\end{align}
where, $r$ determines the statistics of the particle as before. For the massless chargeless gluons we were required to determine the following thermodynamic integrals as,
\begin{align}
    %1
    I_{20}^{g} &= \frac{8\, T^{4}}{15\, \pi^2}, \\
    %2
    I_{21}^{g} &= - \frac{8\, T^{4}}{45\, \pi^2}, \\
    %3
    J_{32}^{g} &= \frac{32\, T^{5}}{225\, \pi^2},
\end{align}
in order to generate the figures in the article.

%---------------------------------------
\section{Various Transport Coefficients}
\label{app:VTC}
%---------------------------------------

In the following, we provide the list of all the second-order transport coefficients appearing Section \ref{ssec:SOTP}. The transport coefficients appearing in Eq.~\eqref{Pi-O2} are,
\begin{align}
    %1
    \beta_0 &= - (2/\beta_\Pi), \label{beta0}\\
    %2
    \beta_{\Pi\Pi} &= - \frac{2}{\beta \beta_\Pi^2} \sum_a \left[\Xi_{30,a}^{(P),+} + \left(\beta_\theta - \frac{\beta}{3} \right) \Xi_{31,a}^{(J), +} - \frac{\beta m^2}{3} \Xi_{11,a}^{(J),+} - \beta \Lambda_\Pi^\Pi - \lambda_{\Pi\Pi\Pi}^{(K)}\right] + \sum_{a} \!\left(\!\frac{2 \beta^{-1}}{\beta_\Pi D_{20}}\!\right) \!\left[\! \left(\frac{J_{20}^{a,-}}{J_{10}^{a,+}}\right) \Xi_{20,a}^{(J),-} \!-\! \Xi_{30,a}^{(J),+}\right] \nonumber\\
    %3
    &\quad- \left(\beta \beta_{\Pi}^2\right)^{-1} \left(\lambda_{\Pi\Pi\Pi}^{(L)} - \lambda_{\Pi\Pi\Pi}^{(J)}/2\right), \\
    %3
    \beta_{\pi\pi} &= \frac{2}{\beta_\pi \beta_\Pi} \sum_a \left[\left(\beta_\theta - \frac{\beta}{3}\right) J_{32}^{a,+} - \left(\frac{\beta m^2}{3}\right) J_{12}^{a,+} - \beta\, \Xi_{42,a}^{(K),+} \right] - \sum_a \left(\frac{2 \beta^{-1}}{\beta_\Pi D_{20}}\right) \left[\left(\frac{J_{20}^{a,-}}{J_{10}^{a,-}}\right) \Xi_{20,a}^{(J),-} - \Xi_{30,a}^{(J),+}\right], \\
    %4
    \psi_{n\mathfrak{a}}^{q'} &= \frac{2}{\beta \beta_\Pi} \sum_{a,q} \hat{\beta}_{qq'} \left[\frac{n_q}{(\varepsilon + P)} \!\left\{\! \left(\beta_\theta \!-\! \frac{\beta}{3}\right) J_{31}^{a,+} \!-\! \left(\frac{\beta m^2}{3}\right) J_{11}^{a,+} \!\right\} \!-\! q_a \!\left\{\! \left(\beta_\theta \!-\! \frac{\beta}{3}\right) J_{21}^{a,-} \!-\! \left(\frac{\beta m^2}{3}\right) J_{01}^{a,-} \!\right\} \right. \nonumber\\
    %5
    &~~ \left.+ \frac{1}{3} \left\{\frac{n_q}{(\varepsilon + P)} \!\left(3 \beta_\theta^{\mathfrak{a}} J_{31}^{a,+} - 5 \beta J_{32}^{a,+} - 3\, \xi_{\theta\mathfrak{a}}^{(a)} J_{21}^{a,-}\right) \!- q_a \!\left(3 \beta_\theta^{\mathfrak{a}} J_{21}^{a,-} - 5 \beta J_{22}^{a,-} - 3\, \xi_{\theta\mathfrak{a}}^{(a)} J_{11}^{a,+} \right)\right\} - \frac{\Lambda_\Pi^{\mathfrak{a}}}{\beta_\Pi} \left\{\frac{n_q\, \Xi_{31,a}^{(J),+}}{(\varepsilon + P)} - q_a\, \Xi_{21,a}^{(J),-} \right\} \right] \nonumber\\
    %6
    &~~- \left(\frac{2}{\beta \beta_\Pi} \right) \sum_a \left[ \hat{\gamma}_{qq'}^{\mathfrak{a}} \left\{\frac{n_q\, \Xi_{31,a}^{(J),+}}{(\varepsilon + P)} - q_a\, \Xi_{21,a}^{(J),-}\right\} + q_a \hat{\beta}_{qq'}\, \Xi_{21,a}^{(J),-} \right],
\end{align}
\begin{align}
%7
    \psi_{nn}^{qq''} &= - \frac{2}{\beta \beta_\Pi} \sum_{a,q'} \hat{\beta}_{q'q''} \left[\xi_{\theta q}^{(a)} \!\!\left\{\! \frac{n_{q'} J_{21}^{a,-}}{(\varepsilon + P)} \!-\! q'_a J_{11}^{a,+} \!\right\} \!-\! \beta_\theta^{q} \!\left\{\! \frac{n_{q'} J_{31}^{a,+}}{(\varepsilon + P)} \!-\! q'_a J_{21}^{a,-} \!\right\} \!-\! \frac{5 n_{q}}{3 (\varepsilon + P)} \!\!\left\{\! \frac{n_{q'} J_{32}^{a,+}}{(\varepsilon + P)} - q'_a J_{22}^{a,-}\right\} \right. \nonumber\\
    %8
    &\quad+ \left.\frac{\Lambda_\Pi^q}{\beta_\Pi} \left\{\frac{n_q \Xi_{31,a}^{(J),+}}{(\varepsilon+P)} - q_a \Xi_{21,a}^{(J),-}\right\} + \left\{\frac{n_q n_{q'}\, \Xi_{41,a}^{(K),+}}{(\varepsilon+P)^2} - \frac{(q_a n_{q'} + q_a' n_{q}) \Xi_{31,a}^{(K),-}}{(\varepsilon+P)} + q_a q_a'\, \Xi_{21,a}^{(K),+}\right\} \right] \nonumber\\
    %9
    &\quad- \sum_{q} \beta_\Pi^{-1} \hat{\beta}_{q'q''} \left(\lambda^{(L)}_{qq'\Pi} - \lambda^{(J)}_{qq'\Pi}/2\right), \\
    %10
    \psi_n^{a'} &= \sum_{a} \! \left(\frac{2 \beta^{-1} J_{20}^{a',-}}{\beta_\Pi D_{20} J_{10}^{a',+}}\right) \!\!\left[\! \left( \frac{J_{20}^{a,-}}{J_{10}^{a,+}} - \frac{D_{20} J_{10}^{a',+}}{J_{20}^{a',-} J_{10}^{a,+}}\!\right) \Xi_{20,a}^{(J),-} \!\!- \Xi_{30,a}^{(J),+} \right], \\
    %11
    \psi_n^{q'} &= - \frac{2 \beta^{-1}}{\beta_\Pi} \sum_{a,q,q'} \hat{\beta}_{qq'} \left[\frac{n_q \Xi_{31,a}^{(J),+}}{(\varepsilon + P)} - q_a \Xi_{21,a}^{(J),-}\right].
\end{align}
The transport coefficients appearing in Eq.~\eqref{nq-O2} are,
\begin{align}
    %1
    \beta_1^{qq'} &= - 2 \beta^{-1} \hat{\beta}_{qq'}, \\
    %2
    \beta_{n\Pi}^{qq'} &= 2 \beta^{-1} \hat{\gamma}_{qq'}^\Pi, \\
    %3
    \beta_{\pi}^{qq'} &= \frac{2}{\beta_\pi} \sum_a \hat{\beta}_{qq'} \left[\frac{n_q J_{32}^{a,+}}{(\varepsilon + P)} - q_a J_{22}^{a,-}\right], \\
    %4
    \beta_{\pi n}^{qq'\!q''} &= - \left(\frac{2}{\beta_\pi}\right) \sum_a \left[\hat{\beta}_{qq'} B_{a}^{qq''} J_{32}^{a,+} + \hat{\gamma}_{qq'}^{q''} \left\{\frac{n_q J_{32}^{a,+}}{(\varepsilon + P)} - q_a J_{22}^{a,-}\right\} \!-\! \hat{\beta}_{qq'} \!\left\{\frac{n_q n_{q''}\, K_{42}^{a,+}}{(\varepsilon + P)^2} - \frac{(q_a n_{q''} + q_a'' n_q) K_{32}^{a, -}}{(\varepsilon + P)} + q_a q_a'' K_{22}^{a,+}\! \right\}\right. \nonumber\\
    %5
    &\quad\left.+\, q_a \beta^{-1} \hat{\beta}_{qq'} \left\{\frac{n_{q''} J_{22}^{a,-}}{(\varepsilon + P)} - q''_a J_{12}^{a,+}\right\} - \hat{\beta}_{qq'} \left\{\frac{n_{q} n_{q''}}{\left(\varepsilon + P\right)^2} K_{42}^{a,+} \!-\! \frac{\left(q_a n_{q''} \!+\! q''_a n_{q}\right)}{\left(\varepsilon + P\right)} K_{32}^{a,-} \!+ q_a q''_a K_{22}^{a,+}\right\}\right], \\
    %6
    \beta_{\Pi n}^{qq'\!q''} &= \left(\frac{2\, \hat{\gamma}_{qq'}^{q''}}{\beta \beta_\Pi}\right) \sum_a \left\{\frac{n_q\, \Xi_{31,a}^{(J),+}}{(\varepsilon + P)} - q_a\, \Xi_{21,a}^{(J),-}\right\} + \left(\frac{2 \hat{\beta}_{qq'}}{3 \beta \beta_\Pi}\right) \sum_{a} \left[\left\{\frac{n_{q''} J_{31}^{a,+}}{(\varepsilon + P)} -q_a'' J_{21}^{a,-}\right\} B_a^{q\Pi} + \frac{5 q_a}{3} \left\{\frac{n_{q''} J_{22}^{a,+}}{(\varepsilon + P)} - q_a'' J_{12}^{a,-}\right\}\right.\nonumber\\
    %7
    &\hspace{2cm} \left.- 6 \left\{\frac{n_q n_{q''}\, \Xi_{41,a}^{(K),+}}{(\varepsilon+P)^2} - \frac{(q_a n_{q''} + q_a'' n_{q}) \Xi_{31,a}^{(K),-}}{(\varepsilon+P)} + q_a q_a''\, \Xi_{21,a}^{(K),+}\right\} + 3\, \Xi_{31,a}^{(J),+} B_a^{qq''} \right], \\
    %8
    \psi_{\pi}^{qq''} &= - \sum_{a} \frac{2 \hat{\beta}_{qq'}}{(\varepsilon+P)} \bigg[\frac{n_q J_{31}^{a,+}}{(\varepsilon+P)} - q_a J_{21}^{a,-} \bigg], \\
    %9
    \psi_{\Pi}^{qq'} &= \frac{2 \hat{\beta}_{qq'}}{\beta \beta_\Pi} \!\sum_{a,q,q'} \! \left[\frac{n_q\, \Xi_{31,a}^{(J),+}}{(\varepsilon + P)} - q_a\, \Xi_{21,a}^{(J),-}\right] + \frac{2 \hat{\beta}_{qq'}}{(\varepsilon+P)} \sum_{a} \bigg[\frac{n_q J_{31}^{a,+}}{(\varepsilon+P)} - q_a J_{21}^{a,-} \bigg].
\end{align}
The transport coefficients appearing in Eq.~\eqref{pi-O2} are,
\begin{align}
    %1
    \beta_2 &= \beta_\pi^{-1}, \\
    %2
    \beta_{\pi\Pi} &= \frac{\beta_\theta}{\beta \beta_\pi} - \frac{7 \beta J_{33}^+}{3 \beta_\pi^2} - \frac{\Lambda_\pi^\Pi J_{42}^+}{\beta_\pi^3} + \frac{\beta}{\beta_\pi^2} \sum_a \Xi_{42,a}^{(K), +} - \frac{1}{\beta_\pi^2} \sum_a \left(\Xi_{32,a}^{(K),+} + \beta \Xi_{42,a}^{(K),+}\right) - \left(\beta \beta_\pi^2\right)^{-1} \left(\lambda_{\pi\pi\Pi}^{(L)} - \lambda_{\pi\pi\Pi}^{(J)}/2\right), \\
    %3
    \varphi_{nn}^{qq'} &= \sum_{a,q'} \frac{\hat{\beta}_{q'q''}}{\beta_\pi} \!\left[\left\{\frac{2 n_{q'}}{\beta (\varepsilon + P)} - \frac{\Lambda_\pi^{q'}}{\beta_\pi}\right\} \left\{\frac{n_q J_{32}^{a,+}}{(\varepsilon + P)} - q_a J_{22}^{a,-}\right\} - 2 \left\{\frac{n_q n_{q'}\, J_{42}^{a,+}}{(\varepsilon + P)^2} - \frac{(q_a n_{q'} + q_a' n_q) J_{32}^{a, -}}{(\varepsilon + P)} + q_a q_a' J_{22}^{a,+}\right\} \right] \nonumber\\
    %4
    &\hspace{2cm}- \left(\beta \beta_\pi^2\right)^{-1} \left(\lambda_{qq'\pi}^{(L)} - \lambda_{qq'\pi}^{(J)}/2\right), \\
    %5
    \varphi_{n\mathfrak{a}}^{q'} &= - \frac{2}{\beta_\pi} \left(2 + \frac{\Lambda_\pi^{\mathfrak{a}}}{\beta_\pi}\right) \sum_{a,q} \hat{\beta}_{qq'}  \left[\frac{n_q J_{32}^{a,+}}{(\varepsilon + P)} - q_a J_{22}^{a,-} \right],
\end{align}
\begin{align}
    %5
    \varphi_{\pi\pi} &= \frac{3 \beta}{\beta_\pi^2} \sum_a \left(K_{33}^{a,+} + \beta K_{43}^{a,+}\right), \\
    %6
    \alpha_{\pi\pi} &= \varphi_{\pi\pi} \pm \left(\beta \beta_\pi^2\right)^{-1} \left(\lambda_{\pi\pi\pi}^{(L)} - \lambda_{\pi\pi\pi}^{(J)}/2\right), \\
    %7
    \alpha_n^{q'} &= \frac{2}{\beta_\pi} \sum_{a,q} \hat{\beta}_{qq'} \left[\frac{n_q\, J_{32}^{a,+}}{(\varepsilon + P)} - q_a J_{22}^{a,-}\right]. \label{alpha_nq}
\end{align}
where we have used the following notations,
\begin{align}
    %1
    \hat{\beta}_{q_{i}^{} q'_{j}} &\equiv \frac{1}{2 \beta_{\rm D}} \varepsilon_{ik\ell} \varepsilon_{jmn} \beta_{q_{k}^{} q'_{m}} \beta_{q_{\ell}^{} q'_{n}}, \\
    %2
    \beta_{\rm D} &\equiv {\rm det} (\boldsymbol{\beta}_{\boldsymbol{qq}'}) = \beta_{BB'} \beta_{QQ'} \beta_{SS'} + 2 \beta_{BQ'} \beta_{BS'} \beta_{QS'} - \beta_{BB'} \beta_{QS'}^2 - \beta_{QQ'} \beta_{BS'}^2 - \beta_{SS'} \beta_{BQ'}^2,
\end{align}
\begin{align}
    %1
    \Lambda_\Pi^\Pi &= - \left(\beta_\theta \beta_\Pi/\beta\right) + \beta_\theta \sum_a \left[- \frac{2 J_{30}^{a,+}}{3\beta} \left(\beta_\theta - \frac{\beta}{3}\right) - \frac{K_{40}^{a,+}}{\beta} \left(\beta_\theta - \frac{\beta}{3}\right)^2 + \left(\!\frac{2 \xi_\theta^{(a)} J_{20}^{a,-}}{3 \beta}\!\right) + \frac{2 \xi_\theta^{(a)} K_{30}^{a,-}}{\beta} \left(\beta_\theta - \frac{\beta}{3}\right)\right. \nonumber\\
    %2
    &~+ \left.\!\frac{2m^2 J_{10}^{a,+}}{3 \beta} \!\left(\!\beta_\theta \!-\! \frac{\beta}{3}\right) \!-\! \frac{2 m^2 J_{10}^{a,+}}{9} \!-\! \frac{K_{20}^{a,+}}{\beta} \! \left\{\!\left(\!\xi_\theta^{(a)}\!\right)^{\!2} \!\!+\! \frac{2\beta m^2}{3} \!\left(\beta_\theta \!-\! \frac{\beta}{3}\right) \!\right\} \!-\! \frac{2m^2 \xi_\theta^{(a)} J_{00}^{a,-}}{3 \beta} \!+\! \frac{2 m^2 \xi_\theta^{(a)} K_{10}^{a,-}}{3} \!+\! \frac{2 m^4 J_{-1,0}^{a,+}}{9} \!-\! \frac{\beta^2 m^4 K_{00}^{a,+}}{9}\right] \nonumber\\
    %3
    &~~+ \left(\beta_{\theta\Pi}/\beta\right) \sum_a \left[2 J_{30}^{a,+} \left(\beta_\theta - \frac{\beta}{3}\right) - 2 \xi_\theta^{(a)} J_{20}^{a,-} + \frac{2\beta m^2 J_{10}^{a,+}}{3} \right] - \sum_a \left(\xi_{\theta\Pi}^{(a)}/\beta\right) \left[2 J_{20}^{a,-} \left(\beta_\theta - \frac{\beta}{3}\right) - 2 \xi_\theta^{(a)} J_{10}^{a,+} + \frac{2 \beta m^2 J_{00}^{a,-}}{3}\right] \nonumber\\
    %4
    &~~+ \sum_{a,q} q_a \xi_\theta^{(a)} \left[\! \frac{K_{30}^{a,-}}{\beta} \!\left(\beta_\theta - \frac{\beta}{3}\right)^2 \!\!-\! \frac{2 \xi_\theta^{(a)} K_{20}^{a,+}}{\beta} \left(\beta_\theta - \frac{\beta}{3}\right) \!+\! \frac{K_{10}^{a,-}}{\beta} \left\{\!\left(\!\xi_\theta^{(a)}\!\right)^{\!2} \!+\! \frac{2\beta m^2}{3} \left(\beta_\theta \!-\! \frac{\beta}{3}\right) \!\right\} \!-\! \frac{2 m^2 \xi_\theta^{(a)} K_{00}^{a,+}}{3} + \frac{\beta m^4 K_{-1,0}^{a,-}}{9}\right],
\end{align}
\begin{align}
    %1
    \Lambda_\Pi^{\mathfrak{a}} &= \beta_\Pi - \beta \sum_a \left[- \frac{2 J_{30}^{a,+}}{3\beta} \left(\beta_\theta - \frac{\beta}{3}\right) - \frac{K_{40}^{a,+}}{\beta} \left(\beta_\theta - \frac{\beta}{3}\right)^2 + \left(\!\frac{2 \xi_\theta^{(a)} J_{20}^{a,-}}{3 \beta}\!\right) + \frac{2 \xi_\theta^{(a)} K_{30}^{a,-}}{\beta} \left(\beta_\theta - \frac{\beta}{3}\right)\right. \nonumber\\
    %2
    &~~+ \left.\!\frac{2m^2 J_{10}^{a,+}}{3 \beta} \!\left(\!\beta_\theta \!-\! \frac{\beta}{3}\right) \!-\! \frac{2 m^2 J_{10}^{a,+}}{9} \!-\! \frac{K_{20}^{a,+}}{\beta} \! \left\{\!\left(\!\xi_\theta^{(a)}\!\right)^{\!2} \!\!+\! \frac{2\beta m^2}{3} \!\left(\beta_\theta \!-\! \frac{\beta}{3}\right) \!\right\} \!-\! \frac{2m^2 \xi_\theta^{(a)} J_{00}^{a,-}}{3 \beta} \!+\! \frac{2 m^2 \xi_\theta^{(a)} K_{10}^{a,-}}{3} \!+\! \frac{2 m^4 J_{-1,0}^{a,+}}{9} \!-\! \frac{\beta^2 m^4 K_{00}^{a,+}}{9} \!\right] \nonumber\\
    %3
    &~~+ \left(\beta_{\theta\mathfrak{a}}/\beta\right) \sum_a \left[2 J_{30}^{a,+} \left(\beta_\theta - \frac{\beta}{3}\right) - 2 \xi_\theta^{(a)} J_{20}^{a,-} + \frac{2\beta m^2 J_{10}^{a,+}}{3} \right] - \sum_{a} \left(\xi_{\theta\mathfrak{a}}^{(a)}/\beta\right) \left[2 J_{20}^{a,-} \left(\beta_\theta - \frac{\beta}{3}\right) - 2 \xi_\theta^{(a)} J_{10}^{a,+} + \frac{2 \beta m^2 J_{00}^{a,-}}{3}\right],
\end{align}
\begin{align}
    %1
    \Lambda_\Pi^q &= - \frac{n_q \beta_\Pi}{\beta (\varepsilon + P)} + \frac{n_q}{(\varepsilon + P)} \sum_a \left[\! - \frac{2 J_{30}^{a,+}}{3\beta} \left(\beta_\theta - \frac{\beta}{3}\right) - \frac{K_{40}^{a,+}}{\beta} \left(\beta_\theta - \frac{\beta}{3}\right)^2 + \left(\!\frac{2 \xi_\theta^{(a)} J_{20}^{a,-}}{3 \beta}\!\right) + \frac{2 \xi_\theta^{(a)} K_{30}^{a,-}}{\beta} \left(\beta_\theta - \frac{\beta}{3}\right)\right. \nonumber\\
    %2
    &~+ \left.\!\frac{2m^2 J_{10}^{a,+}}{3 \beta} \!\left(\!\beta_\theta \!-\! \frac{\beta}{3}\right) \!-\! \frac{2 m^2 J_{10}^{a,+}}{9} \!-\! \frac{K_{20}^{a,+}}{\beta} \! \left\{\!\left(\!\xi_\theta^{(a)}\!\right)^{\!2} \!\!+\! \frac{2\beta m^2}{3} \!\left(\beta_\theta \!-\! \frac{\beta}{3}\right) \!\right\} \!-\! \frac{2m^2 \xi_\theta^{(a)} J_{00}^{a,-}}{3 \beta} \!+\! \frac{2 m^2 \xi_\theta^{(a)} K_{10}^{a,-}}{3} \!+\! \frac{2 m^4 J_{-1,0}^{a,+}}{9} \!-\! \frac{\beta^2 m^4 K_{00}^{a,+}}{9} \!\right] \nonumber\\
    %3
    &~+ \left(\beta_{\theta q}/\beta\right) \sum_a \left[2 J_{30}^{a,+} \left(\beta_\theta - \frac{\beta}{3}\right) - 2 \xi_\theta^{(a)} J_{20}^{a,-} + \frac{2\beta m^2 J_{10}^{a,+}}{3} \right] - \sum_a \left(\xi_{\theta q}^{(a)}/\beta\right) \left[2 J_{20}^{a,-} \left(\beta_\theta - \frac{\beta}{3}\right) - 2 \xi_\theta^{(a)} J_{10}^{a,+} + \frac{2 \beta m^2 J_{00}^{a,-}}{3}\right] \nonumber\\
    %4
    &~+ \sum_a q_a \left[\! \frac{K_{30}^{a,-}}{\beta} \!\left(\beta_\theta - \frac{\beta}{3}\right)^2 \!\!-\! \frac{2 \xi_\theta^{(a)} K_{20}^{a,+}}{\beta} \left(\beta_\theta - \frac{\beta}{3}\right) \!+\! \frac{K_{10}^{a,-}}{\beta} \left\{\!\left(\!\xi_\theta^{(a)}\!\right)^{\!2} \!+\! \frac{2\beta m^2}{3} \left(\beta_\theta \!-\! \frac{\beta}{3}\right) \!\right\} \!-\! \frac{2 m^2 \xi_\theta^{(a)} K_{00}^{a,+}}{3} + \frac{\beta m^4 K_{-1,0}^{a,-}}{9}\right],
\end{align}
\begin{align}
    %1
    \Lambda_{qq'}^\Pi &= \sum_a \left[- \beta_\theta \left\{- q_{a} q'_a K_{21}^{a,+} + \frac{n_q n_{q'} K_{41}^{a,+}}{\left(\beta J_{31}^{a,+}\right)^2} - \frac{2 n_q n_{q'}}{\beta^3 J_{31}^{a,+}} - \frac{1}{J_{31}^{a,+}} \sum_{a'} \left(q_{a'} n_{q'} + q'_{a'} n_{q}\right) J_{20}^{a',-}\right\} \right. \nonumber\\
    %2
    &\left.\qquad+ \xi_\theta^{(a)} \left\{- q_{a} q'_{a} K_{11}^{a,-} + \frac{n_q n_{q'} K_{31}^{a,-}}{(\beta J_{31}^{a,+})^2} - \left(q_{a} n_{q'} + q'_{a} n_q\right) J_{10}^{a,+} \sum_{a'} \left(1/J_{31}^{a',+}\right)\right\} \right],
\end{align}
\begin{align}
    %1
    \Lambda_{qq'}^{\mathfrak{a}} &= \beta \sum_a \left[- q_{a} q'_a K_{21}^{a,+} + \frac{n_q n_{q'} K_{41}^{a,+}}{\left(\beta J_{31}^{a,+}\right)^2} - \frac{2 n_q n_{q'}}{\beta^3 J_{31}^{a,+}} - \frac{1}{J_{31}^{a,+}} \sum_{a'} \left(q_{a'} n_{q'} + q'_{a'} n_{q}\right) J_{20}^{a',-}\right],
\end{align}
\begin{align}
    %1
    \Lambda_{qq'}^{q''} &= \sum_a \left[q''_a \left\{- q_{a} q'_{a} K_{11}^{a,-} + \frac{n_q n_{q'} K_{31}^{a,-}}{(\beta J_{31}^{a,+})^2} - \left(q_{a} n_{q'} + q'_{a} n_q\right) J_{10}^{a,+} \sum_{a'} \left(1/J_{31}^{a',+}\right)\right\}\right. \nonumber\\
    %2
    &\qquad\left.- \frac{n_{q''}}{(\varepsilon + P)} \left\{- q_{a} q'_a K_{21}^{a,+} + \frac{n_q n_{q'} K_{41}^{a,+}}{\left(\beta J_{31}^{a,+}\right)^2} - \frac{2 n_q n_{q'}}{\beta^3 J_{31}^{a,+}} - \frac{1}{J_{31}^{a,+}} \sum_{a'} \left(q_{a'} n_{q'} + q'_{a'} n_{q}\right) J_{20}^{a',-}\right\}\right],
\end{align}
\begin{align}
    %1
    \Lambda_\pi^\Pi &= \sum_a \Big[\beta_\theta \!\left(J_{32}^{a,+} \!- \beta K_{42}^{a,+}\right) + \beta \xi_a K_{32}^{a,-}\Big], \\
    %2
    \Lambda_\pi^{\mathfrak{a}} &= - \beta \sum_a \left(J_{32}^{a',+} \!-\! \beta K_{42}^{a',+}\right), \\
    %3
    \Lambda_\pi^{q} &= \sum_{a} \left[\frac{n_q}{(\varepsilon+P)} \sum_{a'} \! \left(J_{32}^{a',+} \!-\! \beta K_{42}^{a',+}\right) \!+\! \beta q_a K_{32}^{a,-}\right],
\end{align}
\begin{align}
    %1
    \hat{\gamma}_{ij}^X &\equiv \left(2 \beta_{\rm D}\right)^{-1} \left(\varepsilon_{ik\ell} \varepsilon_{jmn} \gamma^X_{q_{k}^{} q'_{m}q_{\ell}^{} q'_{n}} - 2 \hat{\beta}_{q_{i}^{} q'_{j}} \beta_\Gamma^X\right), \\
    %2
    \beta_{\Gamma}^X &\equiv \Gamma^X_{BB'QQ'SS'} + 2 \Gamma^X_{QQ'BS'QS'} - \Gamma^X_{BB'QS'QS'} - \Gamma^X_{QQ'BS'BS'} - \Gamma^X_{SS'BQ'BQ'}, \\
    %3
    \gamma_{q_1q'_1q_2q'_2}^{X} &= \Lambda_{q_1q'_1}^{X} \beta_{q_2q'_2} + \Lambda_{q_2q'_2}^{X} \beta_{q_1q'_1}, \\
    %4
    \Gamma^{X}_{q_1q'_1q_2q'_2q_3q'_3} &= \Lambda_{q_1q'_1}^{X} \beta_{q_2q'_2} \beta_{q_3q'_3} + \beta_{q_1q'_1} \Lambda_{q_2q'_2}^{X} \beta_{q_3q'_3} + \beta_{q_1q'_1} \beta_{q_3q'_3} \Lambda_{q_3q'_3}^{X},
\end{align}
where, $X\in(\Pi, \mathfrak{a}, q)$. The coefficients appearing in Eqs.~\eqref{Pi-evol}-\eqref{pi-evol} are related to those appearing in Eqs.~\eqref{Pi-O2}-\eqref{pi-O2} as,
\begin{align}
    %1
    \tau_{\pi} &= \zeta \beta_0,
    \quad
    \tau_{qq'} = \lambda_{qq'} \beta_1^{qq'},
    \quad
    \tau_{\Pi} = 2 \eta \beta_\pi, \\
    %2
    \delta_{\Pi\Pi} &= \zeta \beta_{\Pi\Pi},
    \quad
    \lambda_{\Pi\pi} = - \zeta \beta_{\pi\pi},
    \quad
    \ell_{\Pi n}^{(q')} = \zeta \psi_{n}^{q'},
    \quad
    \tau_{\Pi n}^{(q')} = \zeta \psi_{n}^{q'},
    \quad
    \lambda_{\Pi n}^{(q,q'')} = \zeta \psi_{n}^{qq''},
    \quad
    \lambda_{\Pi n}^{(a')} = \zeta \psi_{n}^{a'},\\
    %3
    \delta_{nn}^{(q,q')} &=\! \lambda_{qq'} \beta_{n \Pi}^{qq'},
    ~~~
    \ell_{n\Pi}^{(q')} \!=\! - \sum_q \lambda_{qq'} \psi_{\Pi}^{qq'}\!,
    ~~~
    \ell_{n\pi}^{(q')} \!=\! \sum_q \psi_\pi^{qq'}\!,
    ~~~
    \lambda_{n\Pi}^{(q'\!,q'')} \!=\! \sum_{q} \kappa_{qq'} \beta_{\Pi n}^{qq'\!q''}\!\!,
    ~~~
    \lambda_{n\pi}^{(q'\!,q'')} \!=\! \sum_{q} \kappa_{qq'} \beta_{\pi n}^{qq'\!q''}\!\!, \\
    %4
    \delta_{\pi\pi} &= 2 \eta \beta_{\pi\Pi},
    \quad
    \lambda_{\pi n}^{(q,q')} = 2 \eta \varphi_{nn}^{qq'},
    \quad
    \ell_{\pi n}^{(q')} = 2 \eta \alpha_n^{q'},
    \quad
    \tau_{\pi n}^{(q')} = 2 \eta \varphi_{n\mathfrak{a}}^{q'},
    \quad
    \tau_{\pi\pi} = - \frac{28}{5} \eta \varphi_{\pi\pi},
    \quad
    \lambda_{\pi\pi} = 2 \eta \alpha_{\pi\pi}.
\end{align}
Additionally, we have used the following definitions in Eqs.~\eqref{beta0}-\eqref{alpha_nq},
\begin{align}
    %1
    \lambda_{\Pi\Pi\Pi}^{(X)} &=\! \sum_a \bigg[\left(\beta_\theta - \frac{\beta}{3}\right)^{\!3} \!\!X_{40}^{a,+} + \frac{\beta^3 m^6}{27} X_{-20}^{a,+} - \left(\xi_\theta^{(a)}\right)^3 X_{10}^{a,-} + \beta m^2 \left(\beta_\theta - \frac{\beta}{3}\right)^2 X_{20}^{a,+} \nonumber\\
    %2
    &\quad+ \frac{\beta^2 m^4}{9} \left(\beta_\theta - \frac{\beta}{3}\right) X_{00}^{a,+} - \xi_\theta^{(a)} \left(\beta_\theta - \frac{\beta}{3}\right)^2 X_{30}^{a,-} + 3 \left(\xi_\theta^{(a)}\right)^2 \left(\beta_\theta - \frac{\beta}{3}\right) X_{20}^{a,+} \nonumber\\
    %3
    &\quad- \frac{\xi_\theta^{(a)} \beta^2 m^4}{3} X_{-1,0}^{a,-} + \left(\xi_\theta^{(a)}\right)^{\!2} \!\beta m^2 X_{00}^{a,+} - 2\, \xi_\theta^{(a)} \beta m^2 \left(\beta_\theta - \frac{\beta}{3}\right) \!X_{10}^{a,-} \!\bigg], \\
    %4
    \lambda_{\pi\pi\pi}^{(X)} &= 6 \beta^3 X_{43}^{+}, \\
    %5
    \lambda_{qq'\Pi}^{(X)} &= 3 \sum_{a} \Bigg[ \left(\beta_\theta - \frac{\beta}{3}\right) \left\{\frac{n_{q} n_{q'}}{\left(\varepsilon + P\right)^2} X_{41}^{a,+} - \frac{(n_{q} q'_{a} + n_{q'} q_{a})}{(\varepsilon + P)} X_{31}^{a,-} + q'_a q_a X_{21}^{a,+} \right\} \nonumber\\
    %5
    &\quad+ \frac{\beta m^2}{3} \left\{\frac{n_{q} n_{q'}}{\left(\varepsilon + P\right)^2} X_{21}^{a,+} - \frac{(n_{q} q'_{a} + n_{q'} q_{a})}{(\varepsilon + P)} X_{11}^{a,-} + q'_a q_a X_{01}^{a,+} \right\} \nonumber\\
    %6
    &\quad- \xi_\theta^{(a)} \left\{\frac{n_{q} n_{q'}}{\left(\varepsilon + P\right)^2} X_{31}^{a,-} - \frac{(n_{q} q'_{a} + n_{q'} q_{a})}{(\varepsilon + P)} X_{21}^{a,+} + q'_a q_a X_{11}^{a,-} \right\} \Bigg], %\\
\end{align}
\begin{align}
    %7
    \lambda_{\pi\pi\Pi}^{(X)} &= 6  \beta^2 \sum_a \bigg[\left(\beta_\theta - \frac{\beta}{3}\right)^2 X_{52}^{a,+} + \frac{\beta^2 m^4}{9} X_{12}^{a,+} + \left(\xi_\theta^{(a)}\right)^2 X_{32}^{a,-} \nonumber\\
    %8
    &\quad+ \frac{2 \beta m^2}{3} \!\!\left(\beta_\theta \!-\! \frac{\beta}{3}\right) X_{32}^{a,+} - \left(\beta_\theta \!-\! \frac{\beta}{3}\right)\! \xi_\theta^{(a)} X_{42}^{a,-} \!\!-\! \frac{2 \beta m^2 \xi_\theta^{(a)}}{3} X_{22}^{a,-} \bigg], \\
    %9
    \lambda^{(X)}_{qq'\pi} &= 6 \beta \sum_a \left[\!\frac{n_{q} n_{q'}}{\left(\varepsilon + P\right)^2} X_{42}^{a,+} - \frac{\left(q_a n_{q'} + q'_a n_{q}\right)}{\left(\varepsilon + P\right)} X_{32}^{a,-} + q_a q'_a X_{22}^{a,+} 
    \right].
\end{align}

\bibliography{ref}{}

\end{document}